\documentclass{aa}  

\usepackage{natbib}
\bibpunct{(}{)}{;}{a}{}{,} 

\usepackage{array, multirow, graphicx}
\usepackage[dvipsnames]{xcolor}
\usepackage{wasysym}[integrals]
\usepackage{amssymb}
\usepackage{pifont}
\usepackage{siunitx}
\usepackage{booktabs, tabularx}
\usepackage{cellspace, makecell} 
\usepackage{mathtools}
\usepackage{txfonts}

\usepackage[]{hyperref}
\hypersetup{colorlinks,citecolor=blue, linkcolor=blue, urlcolor=blue, breaklinks=true}

\newcounter{magicrownumbers}

\newcommand{\lya}{Ly$\alpha$}

\newcommand{\hi}{\ion{H}{I}}
\newcommand{\civ}{\ion{C}{IV}}
\newcommand{\siv}{\ion{Si}{IV}}
\newcommand{\heii}{\ion{He}{II}}
\newcommand{\abs}[1]{\left | #1 \right |}

\begin{document} 

   \title{Resolving circumgalactic gas flows around a $z\approx3.6$ quasar \\using MUSE and ALMA}


   \author{M. Galbiati\inst{\ref{unimib}}\fnmsep\thanks{\email{marta.galbiati@unimib.it}}
          \and
          A. Pensabene\inst{\ref{unimib},\ref{dawn}, \ref{dtu}}\fnmsep\thanks{\email{anpen@dtu.dk}; second author with equal contribution.}
          \and
          S. Cantalupo\inst{\ref{unimib}}
          \and
          A. Travascio\inst{\ref{unimib}, \ref{inaf_trieste}}
          \and
          G. Pezzulli\inst{\ref{kapt}}
          \and
          R. Decarli\inst{\ref{inaf-bo}}
          \and
          R. Dutta\inst{\ref{iucaa}}
          \and
          S. Muzahid\inst{\ref{iucaa}}
          \and
          J. Schaye\inst{\ref{leiden}}
          \and
          T. Lazeyras\inst{\ref{unimib}}
          \and
          N. Ledos\inst{\ref{unimib}}
          \and
          G. Quadri\inst{\ref{unimib}}
          \and
          W. Wang\inst{\ref{unimib}}
          }
          
\institute{Dipartimento di Fisica ``G. Occhialini'', Universit\`a degli Studi di Milano-Bicocca, Piazza della Scienza 3, I-20126 Milano, Italy\label{unimib}
 \and
 Cosmic Dawn Center (DAWN), Denmark\label{dawn}
 \and
  DTU Space, Technical University of Denmark, Elektrovej 327, DK2800 Kgs. Lyngby, Denmark\label{dtu}
 \and
 INAF--Osservatorio Astronomico di Trieste, Via G.B. Tiepolo, 11, I-34143 Trieste, Italy\label{inaf_trieste}
 \and
 Kapteyn Astronomical Institute, University of Groningen, Landleven 12, NL-9747 AD Groningen, the Netherlands\label{kapt}
 \and
 INAF--Osservatorio di Astrofisica e Scienza dello Spazio, Via Gobetti 93/3, I-40129 Bologna, Italy\label{inaf-bo}
 \and
 IUCAA, Postbag 4, Ganeshkind, Pune 411007, India\label{iucaa}
 \and
 Leiden Observatory, Leiden University, PO Box 9513, 2300 RA Leiden, the Netherlands\label{leiden}
}

   \date{\today}

\abstract{The formation and evolution of galaxies is regulated by the exchange of gas with the surrounding large-scale structures on circum- and intergalactic scales. Yet, little is known about the complex processes shaping the cycle of baryons in and out of galaxies. In this work, we present a multiline study of the gas surrounding a $z\approx3.66$ quasar known to host one of the brightest \lya\ nebulae at high redshift, MUSE Quasar Nebula 04 (MQN04). By combining a high-resolution MUSE detection of non-resonant \heii\ emission with a precise measurement of the redshift of the quasar host via the ALMA CO(4--3) line, we study the kinematics of the cool ionized gas down to $\approx1\rm\,kpc$ from the quasar. The MUSE observations reveal complex clumpy structures as well as diffuse emission extended over $\approx100\,{\rm kpc}$ and blueshifted by $\approx 0-800\,{\rm km\,s^{-1}}$ relative to the quasar systemic redshift, suggesting that the circumgalactic medium is highly asymmetric. The analysis of the \heii/\lya\ line ratio, and the presence of a low-column density ($\approx10^{14.6}~\rm cm^{-2}$) \hi\ absorber along the quasar sightline suggests that MQN04 resides in a highly ionized medium. This is also supported by the gas kinematics, which, except in the most central region, shows consistent velocity shifts across the different tracers, indicative of relatively weak radiative transfer effects. Based on its morphology and kinematics, we conclude that the extended \heii\ emission may arise from merger-driven tidal stripping or inflows of gas illuminated by the quasar radiation. On comoving megaparsec scales, we discover a large concentration ($\delta\approx41$) of star-forming galaxies lying within $\abs{\Delta\varv_{\rm QSO}}\lesssim1000\rm\,km\,s^{-1}$ from the quasar. MQN04 is therefore one of the most overdense environments discovered at this epoch.}


   \maketitle
%

\section{Introduction}
\label{sec:intro}

The gas surrounding galaxies plays a crucial role in regulating their formation and evolution and tracks the cycle and exchange of baryons between galaxies and their environments \citep[e.g. reviews by][]{Tumlinson+2017, Peroux+2020}. Observing the cosmic gas from the filaments of the intergalactic medium (IGM) to the circumgalactic medium (CGM) provides a comprehensive view of the buildup of structures, especially around the epoch of the peak of the cosmic star-formation rate density \citep{Madau+2014}. 

Direct imaging of diffuse gas surrounding galaxies not only reveals its properties but also serves as a powerful diagnostic of the gas-galaxy connection. However, the diffuse nature of this low-density gas makes its detection challenging. Thanks to the advent of integral-field spectrographs such as the Multi Unit Spectroscopic Explorer \citep[MUSE,][]{Bacon+2010} at the Very Large Telescope (VLT), the \hi\ gas has become observable via its \lya\ $\lambda1216$ emission \citep[see the reviews by][]{Cantalupo+2017, Ouchi+2020} down to surface brightness (SB) levels of $\approx10^{-20}-10^{-18}{\rm\,erg\,s^{-1}\,cm^{-2}\,arcsec^{-2}}$. Systematic campaigns around quasar fields at $z\gtrsim3$ have unveiled the ubiquitous presence of \lya\ gaseous nebulae extended over $\gtrsim100\rm\,kpc$ \citep[e.g.][]{Steidel+2000, Cantalupo+2014, Hennawi+2015, Borisova+2016, Arrigoni-Battaia+2019, Fossati+2021} and filamentary structures connecting overdensities of AGNs and star-forming galaxies \citep{Umehata+2019, Bacon+2021, Banerjee+2025, Tornotti+2025}. Although being a powerful tracer of cool diffuse gas, the \lya\ emission suffers from strong radiative transfer effects, hampering the precise determination of the systemic redshift of the quasar host galaxies, and a reliable interpretation of the gas kinematics, which is essential to resolve the gas flows in the CGM. Detecting emission lines other than the \lya\ is therefore crucial. 

At $z\gtrsim3$, the \civ\ $\lambda1548$ and the non-resonant \heii\ $\lambda1640$ transitions can be observed at optical wavelengths \citep[e.g.][]{Villar-Martin+1997, ArrigoniBattaia+2015a, ArrigoniBattaia+2015b, Cai+2016, Cai+2017, Marino+2019, Marques-Chaves+2019}. The combination of such emission lines allows constraining the intrinsic properties of the gas, such as its density, clumpiness, and total mass, as well as the powering mechanism \citep[e.g.][]{Cantalupo+2019, Pezzulli+2019}. Although sparse, the study of such extended nebulae at $z\approx2-4$ points to a picture in which the gas surrounding galaxies is the result of coexisting several processes and components, including multiple intersecting filaments \citep{Herenz+2020}, inflows of gas \citep{Zhang+2023a, Zhang+2023b}, interacting clouds \citep{Jimenez-Andrade+2023}, and galaxies \citep{Sabhlok+2024}. 

A view of the cosmic gas complementary to that provided by rest-frame UV lines of the ionized phases comes from (sub-)millimeter surveys, especially those conducted with the Atacama Large Millimeter Array (ALMA). These provide information on the dust and the cold molecular gas content, and allow a robust determination of the quasar host galaxy systemic redshift \citep[e.g.][]{Ginolfi+2017, Decarli+2019, Li+2021, Emonts+2023, Li+2023, Sulzenauer+2025}. Furthermore, especially in overdense regions, such observations at longer wavelengths have revealed the presence of a population of massive, star-forming and heavily dust-obscured galaxies that were missed in optical data \citep[e.g.][]{Hodge+2013, Miller+2018, Oteo+2018, Jin+2021, Umehata+2015, Umehata+2018, Umehata+2019, Pensabene+2024, Pensabene+2025}, completing the view of the gas-galaxy connection and the effects of the large-scale environment on different galaxy populations. 

In this work, we focus on the MUSE Quasar Nebula 04 (MQN04) surrounding the quasar Q0055-269 at $z\approx 3.66$. This source was originally included in the blind survey for extended \lya\ nebulae conducted by \citet{Borisova+2016}, built on a MUSE observational campaign around 17 radio-quiet quasars at $3.0<z<3.9$ (see Sect.~\ref{ssec:data_muse}). They found that this system hosts one of the brightest \lya-emitting nebulae ($L_{\rm Ly\alpha}=4\times10^{44}\,{\rm erg\,s^{-1}}$) discovered so far at $z\gtrsim3$, extending over $\approx180\,{\rm kpc}$, which also exhibits exceptionally bright \civ\ and \heii\ emission spread across $\gtrsim50\rm\,kpc$ \citep[see also][]{Guo+2020}. In particular, all the tracers revealed the presence of two compact regions located $\approx2.5\arcsec$ away from the quasar. These properties make MQN04 a remarkable laboratory for investigating the structure, ionization, and kinematics of the CGM around quasars. Despite these advances, key aspects of this system remain uncertain. In particular, the lack of a precise systemic redshift for the quasar has so far limited our ability to interpret the kinematics of the surrounding gas. Moreover, the spatial relationship between the extended emission, possible companion galaxies or AGN, and the cold gas phase traced by millimeter observations remains unexplored. To address these questions, we combine new ALMA Band~3 observations with existing MUSE data, providing the first measurement of the quasar systemic redshift and enabling a direct comparison between the ionized and molecular gas phases. Together with new high-resolution MUSE Narrow-Field Mode (NFM) observations, this dataset allows us to (i) reliably trace the gas kinematics of the nebula through non-resonant \heii\ emission, (ii) search for companion galaxies or AGN within the inner few kiloparsecs, and (iii) investigate the morphology and clumpiness of the CGM on small scales. These complementary observations offer an unprecedented opportunity to understand the interplay between quasar activity, gas flows, and the assembly of massive galaxies at high redshift.

The structure of the paper is as follows: we describe the details and reduction processes of the data in Sect.~\ref{sec:data}. In Sect.~\ref{sec:galaxies}, we describe the identification of galaxies in the field. In Sect.~\ref{sec:nebule}, we present the detection and properties of the extended gaseous nebulae, and a characterization of the large-scale galaxy environment. In Sect.~\ref{sec:disc}, we discuss possible scenarios that can be invoked to explain the observed emission. Finally, we summarize the key results of this work and draw our conclusions in Sect.~\ref{sec:conclusions}. Throughout this work, we adopt a standard $\Lambda$ cold dark matter cosmology with $H_0=67.7\,{\rm km\,s^{-1}\,Mpc^{-1}}$, $\Omega_{\rm m}=0.310$, and $\Omega_\Lambda=1-\Omega_{\rm m}$ from \citet{PlanckColl+2020}, make use of AB magnitude, report distances
in physical units, and quote $3\sigma$ limits unless stated otherwise. 

\section{Observations and data reduction}
\label{sec:data}

\subsection{MUSE observations}
\label{ssec:data_muse}

We analyze MUSE observations\footnote{The data have been released in 2017 and are publicly available on the ESO archive at \url{https://doi.eso.org/10.18727/archive/42}} centered on the quasar Q0055-269 at $z\approxeq3.66$. These have been collected as part of the MUSE Guaranteed Time Observations (GTO; program IDs 094.A-0131(B) and 096.A-0222(A), PI J. Schaye) and are included in the MUSE Quasar-field Blind Emitters Survey \citep[MUSEQuBES;][]{Muzahid+2020}. The data consists of a combination of 14 exposures taken in WFM-NOAO-N mode and resulting in an integrated exposure time of 10h on source. To mitigate the impact of bad pixels and cosmic rays, each exposure was shifted by a small offset and rotated by $\ang{90}$ relative to the previous one. The MUSE field-of-view (FoV) spans $1\arcmin\times1\arcmin$ with a sampling of $0.2\arcsec$ per pixel and covers the optical wavelength range 4800-9300 {\rm\AA} with a sampling of 1.25 $\rm\AA$ and a spectral resolution ranging from $R\approx1800~\text{to}~\approx3600$. We refer to \citet{Muzahid+2021} for details about the data reduction, while only a brief description of the key steps is given here. First, the raw data are processed using the ESO MUSE pipeline \citep[][version 1.6]{Weilbacher+2020} to perform bias subtraction, flat fielding, twilight corrections, wavelength calibration, and corrections for telluric absorption. To improve the quality of the final datacube, the products of this stage of data reduction are post-processed using the {\sc CubExtractor} package ({\sc CubEx} hereafter, version 1.6, \citealp[see][]{deBeer+2023} and Cantalupo et al., in prep.). The subroutines {\sc CubeFix} and {\sc CubeSharp} are employed to perform self-calibration and flux-conserving sky subtraction, respectively \citep[see e.g.][]{Borisova+2016, Bacon+2017, Marino+2018, Cantalupo+2019}. Finally, the individual exposures are combined into a fully reduced datacube using the {\sc CubeCombine} tool. We calibrated the astrometry of the datacube using stars from \citet{Gaia+2023} In the final datacubes, we obtain a \lya\ $1\sigma$ surface brightness limit of $\approx1.5\times10^{-19}\,{\rm erg\,s^{-1}cm^{-2}arcsec^{-2}}$ (corresponding to $\approx4.2\times10^{-21}\,{\rm erg\,s^{-1}cm^{-2}pixels^{-2}}$) in a single layer of $1.25$\AA\ within circular apertures of radius $r=1\arcsec$ (corresponding to 5~pixels).

We also performed MUSE NFM-AO-N observations of the central quasar (PID 109.232M, PI: A. Travascio). The key steps of the data reduction are the same as those detailed for the WFM observations above, except for the treatment of the sky. While in the case of WFM observations, the sky frame was generated directly from the OBJECT exposures, for the NFM data we created the sky frame from the pixel tables of additional exposures taken in empty regions of the sky. The observing sequence consists of 250s of sky exposures every 2300s of OBJECT ones. The reduced and combined data result in an integrated on-source exposure time of 2.3h, spanning a $7.4\arcsec\times7.4\arcsec$ FoV with a sampling of $0.025\arcsec$ per pixel. The spectral coverage is the same as for the WFM observations, except for the range $5800-6000\rm\,\AA$ which is masked due to the contamination of the \ion{Na}{D} lines of the notch filter required for the laser guide star in AO mode. Due to the absence of bright stars in the small FoV, we calibrated the astrometry of the combined datacube by aligning the centroid of the QSO emission to the ALMA detection of its CO(4--3) emission line (described in Sect.~\ref{ssec:galaxies_alma}). For the NFM data, we obtain in circular apertures of radius $r=0.125\arcsec$ (corresponding to 5~pixels) a \lya\ $1\sigma$ SB limit in a single layer of $\approx1.4\times10^{-17}\,{\rm erg\,s^{-1}cm^{-2}arcsec^{-2}}$ (corresponding to $\approx8.7\times10^{-21}\,{\rm erg\,s^{-1}cm^{-2}pixels^{-2}}$).
Finally, we characterize the point spread function (PSF) of the MUSE observations as described in Sect.~\ref{ssec:nebule_cubex}. The effective PSF full width at half maximum (FWHM) estimated from the central quasar is $\approx0\rlap{.}{\arcsec}9$ and $\approx0\rlap{.}{\arcsec}12$ in the WFM and NFM data, respectively.

\subsection{ALMA observations}
\label{ssec:data_alma}

We employ data from ALMA Cycle 11 program 2024.1.00499.S (PI: A. Pensabene). The observations employed ALMA band 3 designed to target the rotational transition of carbon monoxide CO(4--3) around the quasar redshift of $z\approx 3.66$ (rest-frame frequency $461.041\,{\rm GHz}$) as well as the underlying 3 mm dust continuum (rest-frame wavelengths $650\,{\rm \mu m}$) with medium angular resolution of $\approx 0\rlap{.}{\arcsec}35$ in the quasar field. The half power beam width of the 12m ALMA antenna is $58\rlap{.}{\arcsec}8$ at the reference frequency of $99.00\,{\rm GHz}$. The frequency setup consists of two adjacent 4-bit $0.938\,{\rm GHz}$-wide spectral windows (SPWs) in the lower-side band (LSB) covering the CO(4--3) line within $\approx \pm 2780\,{\rm km\,s^{-1}}$ around the quasar systemic redshift, and two 2-bit $1.875\,{\rm GHz}$-wide SPWs placed in the upper-side band (USB) for continuum detection. The native spectral resolution  of the acquired data is $\approxeq 3.9\,{\rm MHz\,channel^{-1}}$ ($\approxeq 11.8\,{\rm km\,s^{-1}\,channel^{-1}}$ at $99.00\,{\rm GHz}$). Observations were taken in five execution blocks during the period 2-5 April 2025 employing a total on-source exposure time of $\approx 4.07\,{\rm h}$, and utilizing a number of ALMA main array antennas ranging from 42 to 47 with baselines within $15.1-2516.9\,{\rm m}$ yielding a naturally-weighted synthesized beam FWHM of $0\rlap{.}{\arcsec}54\times0\rlap{.}{\arcsec}40$. The weather conditions during the executions provided a mean precipitable water vapor between 0.9 and $1.9\,{\rm mm}$ achieving a nominal sensitivity of $48\,{\rm \mu Jy\,beam^{-1}}$ over a bandwidth of $33.190\,{\rm MHz}$ ($\approxeq 100\,{\rm km\,s^{-1}}$) for line observations and $6\,{\rm \mu Jy\,beam^{-1}}$ over the aggregate continuum bandwidth of $4.363\,{\rm GHz}$.

We calibrated the data using the Common Astronomy Software Application (CASA, \citealt{McMullin+2007, Hunter+2023}) by running the calibration pipeline \texttt{scriptForPI} provided along with the measurement using CASA version 6.6.1. No self-calibration was performed on the dataset. We imaged the ALMA band 3 visibilities by using the CASA \texttt{tclean} task. We first obtained a "dirty" datacube (\texttt{specmode="cube"}) with $25\,{\rm km\,s^{-1}}$ channel width, and a continuum image (\texttt{specmode="mfs"}) by Fourier transforming raw visibilities (\texttt{niter=0}) adopting a natural weighting scheme to maximize the sensitivity per beam. We set the pixel scale to $0\rlap{.}{\arcsec}08$ to achieve Nyquist sampling of the beam minor axis with at least $5$ pixels. Additionally, in order to improve the surface brightness sensitivity for faint extended emission, we obtained a "dirty" continuum image and cube applying a Gaussian taper in the $uv$ plane with ${\rm FWHM}_{uv}\approxeq182\,{\rm k\lambda}$ and setting the pixel scale to $0\rlap{.}{\arcsec}2$, yielding a synthesized beam FWHM of $1\rlap{.}{\arcsec}36\times1\rlap{.}{\arcsec}27$. Finally, we obtained "cleaned" data by running the  \texttt{tclean} task down to a signal-to-noise threshold of $\texttt{nsigma=1.0}$ placing masks around the observable emission of the sources (see Sect.~\ref{sec:galaxies}). 

\subsection{Ancillary archival datasets}

We complement our observations with archival imaging and spectroscopic data of the quasar and its surrounding field, as described below.

\subsubsection{HST imaging}
\label{ssec:data_hst}

The {\it Hubble} Space Telescope (HST) observed the field around the Q0055-269 on 2024 June 25 (program ID: 17483; PI: R. Dutta) using the F160W filter on the Wide Field Camera 3 (WFC3) infrared (IR) detector with a total integration time of $1809\,{\rm s}$ split into three frames with dithering offsets as in the {\tt WFC3-IR-DITHER-LINE-3PT} template. The data reduction was performed using the standard HST pipeline, in particular the {\tt DrizzlePac} version 3.9.1 \citep{Hoffmann+2021, Fruchter+2010}. The pixel scale of the final image is $0\rlap{.}{\arcsec}128\,{\rm pix^{-1}}$. At the redshift of the structure, $z\approx3.66$, the pivot wavelength of the filter ($\lambda_{\rm ref}=15370\,\AA$) samples the rest-frame UV ($329.8\,{\rm nm}$) emission. The data described here can be obtained from the MAST archive at \url{http://dx.doi.org/10.17909/7rps-jf20}. Finally, we obtained a background-subtracted HST/WFC3 F160W image by using {\tt photutils} Python package \citep{Bradley+2023} with sigma-clipping statistics to remove the sources, and {\tt SExtractionBackground} as an estimator of the background. The astrometry of the final product is registered to \citet{Gaia+2023}.

\subsubsection{UVES spectrum of the quasar}
\label{ssec:data_uves}

High-resolution spectroscopy of the quasar Q0055-269 taken with the Ultraviolet and Visual Echelle Spectrograph \citep[UVES, ][]{Dekker+2000} at the VLT was already publicly available\footnote{\url{https://github.com/MTMurphy77/UVES_SQUAD_DR1}, \url{https://doi.org/10.5281/zenodo.1345974}} from the first data release of the Spectral Quasar Absorption Database (SQUAD) project \citep{Murphy+2019}. We retrieved the coadded and continuum-normalized UVES spectrum of the quasar. This consists of a combination of 24 exposures (programme IDs 092.A-0011(A) and 65.O-0296(A), PI: J. Schaye, S. D’Odorico) resulting in a total exposure time $t_{\rm exp}=89245~{\rm s}$ and covering the wavelength range $3139.73-9766.69~{\rm\AA}$ with a resolution of $R\approx45,000$. This data provides a complementary view of the quasar environment in absorption, which we later compare with the results obtained in emission from MUSE.

\begin{figure*}[!t]
    \centering
    \includegraphics[width=1\linewidth]{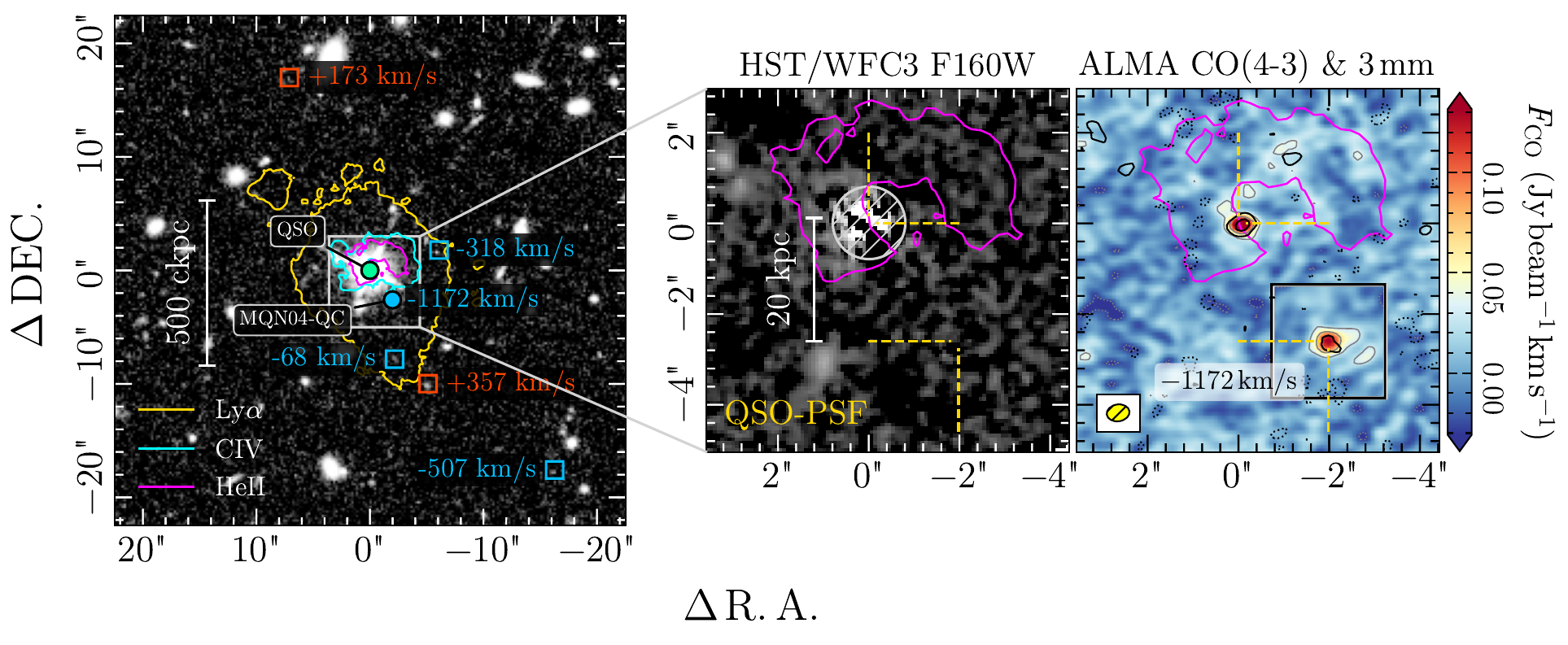}
    \caption{Spatial distribution of galaxies in the MQN04 field detected within $\abs{\Delta\varv}\le1000\rm\,km\,s^{-1}$, the ALMA-identified MQN04-QC galaxy at $\Delta\varv \approx -1172\,{\rm km\,s^{-1}}$ of the quasar systemic redshift, and the \lya, \civ, and \heii-emitting gas. {\it Left panel}: HST/F160W image. The star-forming galaxies detected in MUSE data are indicated by red and blue squares, depending on their line-of-sight velocity relative to the quasar's systemic redshift as determined from the CO(4--3) line emission. The quasar host and the MQN04-QC companion galaxy are shown as circles. The contours indicate the detected extent of the \lya\ (gold), \civ\ (cyan), and \heii\ (magenta) nebulae (see Sect.~\ref{sec:nebule}). {\it Central panel}: QSO PSF-subtracted cutout of the HST/F160W image. The dashed lines mark the position of the quasar host and MQN04-QC galaxy as identified from ALMA observations. The region masked for the PSF modeling is also shown as a hatched circle (see Sect.~\ref{ssec:galaxies_alma}). {\it Right panel}: ALMA CO(4--3) line-velocity integrated map and 3 mm continuum. The gray (CO(4--3) line emission) and black (3 mm continuum) contours correspond to $[-2,2.5^n]\sigma$ (dotted and solid for the negative and positive levels, respectively), where $n\ge1$ is an integer, and $\sigma$ is the RMS noise. The ALMA synthesized beam is shown as a yellow ellipse in the bottom left corner.}
    \label{fig:pin-point-gals}
\end{figure*}

\section{Identification of galaxies in the field}
\label{sec:galaxies}

Leveraging the broad wavelength coverage of our data, we searched within the field to identify different galaxy populations and explore how they relate to the gas observed in emission.  

\subsection{ALMA-detected galaxies}
\label{ssec:galaxies_alma}

We inspected ALMA datasets and identified the CO(4--3) line and 3 mm continuum emission of the quasar host galaxy. Additionally, we detected two other sources (hereafter named MQN04-QC and AG1) in the field showing emission lines at $\approx -1170\,{\rm km\,s^{-1}}$ and $\approx -2200\,{\rm km\,s^{-1}}$ relative to the quasar systemic redshift, and marginally detected in the 3 mm continuum. In particular, we note that MQN04-QC is detected at 3 mm with $F_{\rm 3mm}=19\pm7\,{\rm \mu Jy}$, corresponding to a signal-to-noise ratio of ${\rm S/N\approx2.5}$. The spatial coincidence of the continuum emission peak with that of the CO(4--3) line emission strengthens the reliability of this detection (see Fig~\ref{fig:pin-point-gals}).

We searched for counterparts of these two ALMA-detected galaxies in the HST/WFC3 F160W image and the MUSE datacube. We detected stellar continuum and a line emission in the spectrum of AG1, which we identified as [\ion{O}{II}] at $z\approx1.314$ using \texttt{Marz} tool \citep{Hinton+2016}. Therefore, the emission line detected in ALMA corresponds to CO(2--1). MQN04-QC remains instead undetected in MUSE observations. We therefore searched for rest-frame UV emission in the HST/WFC3 F160W image at the location of MQN04-QC. To do so, we first subtracted the PSF of the QSO from the image. To model the PSF, we used one star in the field with similar observed flux to the quasar as a template, then performed a fit using MCMC ensemble sampler {\tt emcee} \citep{Foreman+2013} using 30 walkers, and 500 steps discarding the first 100 as a burning phase. To optimize the subtraction of the PSF spikes and wings, we masked the quasar emission within the central $0\rlap{.}{\arcsec}8$. In addition, we exclude outer regions beyond $2\rlap{.}{\arcsec}8$ where the presence of additional sources may introduce biases in the fit. We used as free parameters the flux rescaling factor and the position of the PSF centroid, assuming uniform priors. The resulting PSF-subtracted image is shown in Fig.~\ref{fig:pin-point-gals}. We did not detect any counterparts of MQN04-QC or any additional sources\footnote{The faint diffuse emission that in Fig.~\ref{fig:pin-point-gals} is visible eastward to the quasar is likely associated with residuals of the PSF subtraction due to the simplified model adopted.}. The absence of any detected rest-frame UV emission associated with this galaxy could be explained by the presence of a large amount of dust obscuring its stellar light, as commonly observed in high-redshift submillimeter galaxies \citep[SMGs, see, e.g.][]{Casey+2014}. This may suggest that this source is a dust-obscured quasar companion at $z\approx3.65$. Given the lack of additional redshift information, the observed line may be interpreted either as a lower-J ($J_{\rm up}\le3$) or $J_{\rm up}\ge5$ CO lines, the frequency of which enter in the LSB of the ALMA observations from redshift $z\lesssim2.5$ and $z\gtrsim4.8$, respectively. However, based on the CO luminosity functions from \citet{Decarli+2019c, Boogaard+2023}, the expected number of galaxy interlopers in the volumes probed by the current ALMA observations is $<0.2$. For the subsequent analysis, we therefore assume MQN04-QC to be located at $z=3.65$.

We analyze the CO(4--3) line emission of the quasar host and MQN04-QC by extracting spectra from the naturally-weighted ALMA band 3 datacube. To this purpose, we applied the 2-$\sigma$ photometry method \citep[e.g.][]{Bethermin+2020, Pensabene+2024} which enables recovering the line flux with an optimal S/N while preventing significant flux losses: ({\it i}) we extracted a spectrum from the source peak pixel as determined by the 3 mm dust continuum image, we corrected for the primary beam (PB) response, and fit the spectrum with a Gaussian component for the line profile and a constant for the continuum. ({\it ii}) We obtain the line-velocity integrated map (0th moment) by summing together all the channels within $\pm2\sigma$ from the fitted line centroid of the initial spectrum. ({\it iii}) We extracted a new spectrum by summing the signal from all the corresponding pixels in the 0th moment map enclosed within the $S/N \ge2$ isophote. We converted the spectrum in flux density by dividing it by the synthesized beam area (in pixel units), and computed the associated uncertainties by multiplying the peak flux uncertainty (in ${\rm mJy\,beam^{-1}}$) by the square root of the independent beam elements over the extracting region. ({\it iv}) We repeated the line fit on this new spectrum as described above, and we iterated the procedure until convergence. Hence, we performed an accurate fit of the resulting line profile by minimizing the residuals between the data and the model using the MCMC ensemble sampler {\tt emcee} Python package \citep{Foreman+2013} with 100 walkers, 5000 steps, and discarding the first 1000 as a burning phase. We adopted Gaussian uncertainties in the definition of the likelihood and flat priors on the free parameters based on the last fitting iteration as described above. Finally, we derived the line luminosity as \citep[e.g.][]{Solomon+1997}:
$L_{\rm CO}\,[L_{\astrosun}]=1.04\times10^{-3}\,F_{\rm CO}\,\nu_{\rm obs}\,D_{\rm L}^{2}$, $
L^\prime_{\rm CO}\,[{\rm K\,km\,s^{-1}pc^{2}}]=3.25\times10^{7}\,F_{\rm CO}\,{D_{\rm L}^2\,}{(1+z)^{-3}\,\nu_{\rm obs}^{-2}}$,
where $F_{\rm CO}$ is the velocity-integrated line flux in units of ${\rm Jy\, km\,s^{-1}}$, $\nu_{\rm obs}=\nu_{\rm rest}/(1+z)$ is the observed central frequency of the line in GHz at the redshift $z$, and $D_{\rm L}(z)$ is the luminosity distance in Mpc. The two luminosity measurements are related via $L_{\rm CO}=3\times10^{-11}\,\nu_{\rm rest}^{3}\,L'_{\rm CO}$. 

\begin{figure*}[!ht]
    \centering
    \includegraphics[width=0.8\linewidth]{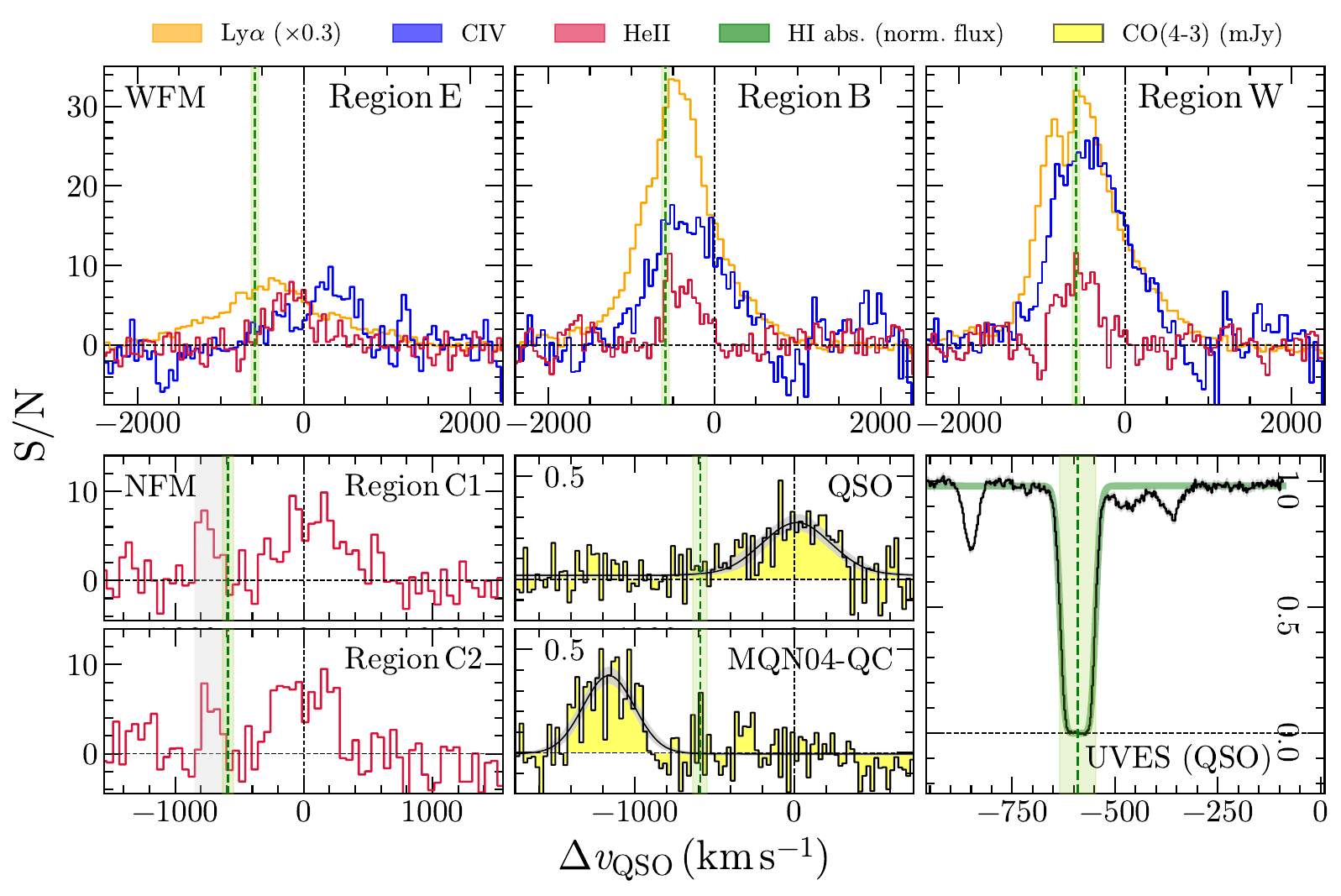}
    \caption{Signal-to-noise spectra extracted from the regions identified in the line-emitting nebulae. {\it Top panels}: \lya\ (orange), \civ\ (blue), and \heii\ (red) emission lines extracted from the MUSE-WFM data within the regions E (left), B (middle), and W (right). The \lya\ S/N is re-scaled by $\times0.3$ for visualization purposes. {\it Bottom-left panels}: \heii\ emission line S/N spectra extracted from the MUSE-NFM data within the clumps C1 (top) and C2 (bottom). The shaded band marks spectral regions contaminated by skylines. {\it Bottom-center panels}: CO(4--3) emission line profiles of the quasar host (top) and the MQN04-QC galaxy (bottom). The data are shown in yellow, the single Gaussian best-fit curve is in black with $1\sigma$ uncertainty shaded in gray. The flux is in units of $\rm mJy$. {\it Bottom-right panel}: \hi\ absorption line detected in the (continuum-normalized) UVES spectrum of the quasar. The green line indicates the best-fit Voigt profile. In all the panels, the line-of-sight velocity is relative to the systemic redshift of the quasar host galaxy as determined from its CO(4--3) emission line. The green dashed line and shaded area indicate the centroid and the FWHM of the \hi\ absorption line, respectively.}
    \label{fig:spectra}
\end{figure*}

In addition to line measurements, we recovered the faint (observed-frame) 3 mm continuum emission of the sources by applying a single-beam photometry measuring the continuum source flux peak in the $\approx1\rlap{.}{\arcsec}3$ tapered image. In Fig.~\ref{fig:pin-point-gals}, we show the velocity-integrated CO(4--3) line maps of the quasar host and the foreground companion galaxy MQN04-QC, as well as the HST/WFC3 image in the F160W filter. In Fig.\ref{fig:spectra} we show the extracted CO line profile of the sources along with the best fit model. We report our measurements in Table~\ref{tbl:sou_flux_prop}.

We estimate an order-of-magnitude indication of the dust and molecular gas mass of MQN04-QC and the quasar host galaxy from the observed 3-mm flux density and the CO(4--3) line luminosity. For this purpose, we model the dust emission with a single-temperature modified blackbody. In the case of MQN04-QC, we adopt a dust emissivity index of $1.8$, and assume a dust temperature of $T_{\rm dust}=32~{\rm K}$, as observed in sub-mm galaxies (SMGs) at $z\sim3$ \citep[e.g.,][]{Dudzeviciute+2020}. Under these hypotheses, we obtain a dust mass ranging within $M_{\rm dust}\sim (0.8-1.6)\times10^8\,M_\odot$, taking into account the 3-mm flux density uncertainties. We then estimate the gas mass by assuming a CO(4--3)-to-CO(1--0) luminosity conversion factor of $r_{41}=0.46$ from \citet{CarilliWalter2013}, and a CO(1-0) luminosity-to-H$_2$ molecular gas mass conversion factor of $\alpha_{\rm CO}=1.7\,M_{\odot}\,({\rm K\,km\,s^{-1}\,pc^2})^{-1}$ measured in a sample of local ultra-luminous infrared galaxies (\citealt{MontoyaArroyave+2023}, see also \citealt{Bolatto+2013} for a review). With these assumptions we obtain $M_{\rm gas}=\alpha_{\rm CO}\,r_{41}^{-1}\,L'_{\rm CO}\sim2\times10^{10}\,M_\odot$, and therefore a gas-to-dust ratio of $\delta_{\rm dg} \sim 95-190$. These estimates are in broad agreement with typical properties of SMGs at $z=2-4$ \citep[see,e.g.][]{Swinbank+2014, Dudzeviciute+2020, Pantoni+2021}. With the same approach, we obtain estimates of the gas and dust mass for the quasar host galaxy. In this case, according to expectations for high-redshift quasar hosts \citep[see, e.g.,][]{Downes+1998, Beelen+2006, CarilliWalter2013, Leipski+2013, Leipski+2014}, we assume $T_{\rm dust}=45\,{\rm K}$, $r_{41}= 0.87$, $\alpha_{\rm CO}=0.8\,M_{\odot}({\rm K\,km\,s^{-1}\,pc^2})^{-1}$. We obtain $M_{\rm dust}\sim (1.3-2)\times10^8\,M_\odot$, $M_{\rm gas}\sim5\times10^{9}\,M_\odot$, and a gas-to-dust ratio of $\delta_{\rm dg} \sim 20-40$. Finally, under the assumption of dispersion dominated systems, we derive a rough estimate of the virial dynamical mass enclosed within a radius $R$ of the two galaxies using the equation $M_{\rm dyn}(<R)\approx(3/2)\,\sigma^2\,R/G$ \citep[e.g.,][]{Decarli+2018, Pensabene+2025}. With $\sigma$ as the CO linewidth and $R$ as half of the beam size $\approx0\rlap{.}{\arcsec}25\approx 1.8\,{\rm kpc}$, considered an upper limit for the unresolved galaxy size, we obtain $M_{\rm dyn}(<1.8\,{\rm kpc}) = (3.1\pm1.2)\times10^{10}\,M_{\odot}$ and $(1.6\pm0.3)\times10^{10}\,M_{\odot}$ for the quasar host galaxy and MQN04-QC, respectively. These values are in agreement with the gas masses inferred from the CO line luminosity. In particular, the gas mass of MQN04-QC appears to be comparable to its dynamical mass, therefore suggesting that this galaxy is very rich in molecular gas. However, we stress that the evaluation of the various mass budgets provided above relies on assumptions that cannot be verified with the limited available information, and therefore should only be taken as order-of-magnitude estimates.

\begin{table}[!t]
\def\arraystretch{1.15}
\caption{Line and continuum flux measurements, luminosity estimates, and derived quantities of the ALMA-identified CO emitters.}  
\label{tbl:sou_flux_prop}    
\centering
\resizebox{0.9\hsize}{!}{
\begin{tabular}{l c c}
\toprule
\toprule
 & Q0055-269 & MQN04-QC$\,^{(\dagger)}$  \\
\cmidrule(lr){1-3}
R.A. (ICRS) & 00:57:58.030   & 00:57:57.882  \\
Dec. (ICRS) & -26:43:14.716 & -26:43:17.314 \\
\bottomrule
Line & CO(4--3) & CO(4--3)\\
\cmidrule(lr){1-3}
$z_{\rm CO}$ & $3.6639^{+0.0006}_{-0.0007}$ & $3.6457^{+0.0003}_{-0.0003}$ \\
$\Delta\varv_{\rm QSO}$ (km s$^{-1}$) &  $-$ & $-1172^{+47}_{-48}$ \\
$F_{\rm CO}$ (Jy km s$^{-1}$) & $0.14^{+0.02}_{-0.02}$ & $0.15^{+0.02}_{-0.02}$ \\
FWHM (km s$^{-1}$) & $512^{+104}_{-87}$ & $376^{+44}_{-40}$ \\
$L_{\rm CO}$ ($10^8 L_\odot$) & $0.16^{+0.03}_{-0.02}$ & $0.17^{+0.02}_{-0.02}$ \\
$L'_{\rm CO}$ ($10^{10}$ K km s$^{-1}$ pc$^2$) & $0.51^{+0.09}_{-0.08}$ & $0.55^{+0.07}_{-0.06}$ \\
\bottomrule
Observed-frame continuum & 3 mm & 3 mm \\
\cmidrule(lr){1-3}
$F_{\rm 3mm}$ (${\rm \mu Jy}$) & $41\pm7$ & $19\pm7$ \\
\bottomrule
\end{tabular}
}
\tablefoot{$^{(\dagger)}$No additional redshift information is available for MQN01-QC. The line redshift and luminosities are derived interpreting the observed line as the CO(4--3) transition.}
\end{table}

\subsection{MUSE continuum-detected galaxies}
\label{ssec:galaxies_muse}

Additionally, we perform a blind search for continuum-detected star-forming galaxies in the MUSE-WFM datacube following the approach described by \citet{Galbiati+2025}. Briefly, we collapsed the datacube along the wavelength axis, producing a white-light image used for the detection. We then extracted the continuum-emitting sources by running {\tt SExtractor} \citep[][version 2.24.2]{Bertin+1996} with {\sc threshold}=1.5 weighted on the local RMS background level to make it uniform across the FoV, using {\sc minarea}=3 pixels, and default deblending parameters. The spectra of each source were then extracted within an aperture of radius $r=2.5\times R_{\rm Kron}$, where $R_{\rm Kron}$ is the Kron radius, chosen to collect the total flux of the sources. Finally, we determined the spectroscopic redshifts using the {\tt Marz} tool\footnote{We used the high-resolution templates of high-redshift galaxies included in the M. Fossati fork available at \url{https://matteofox.github.io/Marz/}.} based on \lya~$\lambda1216$\AA\ emission line and, if visible, looking for \ion{Si}{II}~$\lambda$1260,~$\lambda$1526\AA, \ion{O}{I}+\ion{Si}{II}~$\lambda$1303\AA, \ion{C}{II}~$\lambda$1334\AA, \ion{Si}{IV} $\lambda$1393, 1402\AA, and \ion{C}{IV}~$\lambda$1548, 1550\AA. With this procedure, we identified five galaxies within $\abs{\Delta\varv_{\rm QSO}}\le1000~\rm{km\,s^{-1}}$ relative to the central quasar\footnote{For the majority of these galaxies, the redshift is estimated using the \lya\ emission line. However, due to radiative transfer effects, we expect the \lya\ line to be offset from the systemic redshift of the galaxies. To test whether these effects significantly impact our estimate of spectroscopic redshifts, we employed the empirical relation derived by \citet{Verhamme+2018}, which is based on the shape of the emission line. We measured an average offset of $\approx650\rm\,km\,s^{-1}$. Even accounting for such offsets, all the five galaxies would still lie within $\abs{\Delta\varv_{\rm QSO}}\le1000\rm\,km\,s^{-1}$ from the central quasar.} and with $r$-band magnitudes $m_{\rm r}\lesssim27\,{\rm mag}$ (see Table~\ref{tbl:muse_galaxies}). In Fig.~\ref{fig:pin-point-gals}, we show the location of all the ALMA- and MUSE-selected galaxies, together with the maximum extent of the detected nebular emission (see Sect.~\ref{sec:nebule}).
We note that additional analysis is required to obtain a complete census of galaxies in the field, since emission-line dominated galaxies, such as fainter \lya\ emitters (LAEs), could be missed by our current search for continuum-bright sources. However, the presence of the bright and extended \lya\ emission in the FoV prevents an unambiguous identification of LAEs around the quasar in the absence of continuum-detected counterparts and spectral features other than their \lya\ line.

\section{Detection of \lya, \civ, \heii\ extended emission}
\label{sec:nebule}

\subsection{The gas in emission: extraction of the nebulae}
\label{ssec:nebule_cubex}

We searched the MUSE datacube for extended \lya\ $\lambda1216$\AA, \civ\ $\lambda1548$\AA, and \heii\ $\lambda1640$\AA\ emission around the central quasar in the WFM observations as well as, with higher resolution, for the central part of the extended \heii\ emission in the NFM ones\footnote{In the case of NFM data, we limit the extraction to the \heii\ line emission only, since in this datacube the \lya\ emission line partially falls in the masked AO gap. In addition, the presence of a bright skyline prevents a reliable extraction of the \civ\ line emission. Indeed, compared to WFM observations, the characterization of sky emission in the NFM data is less accurate due to the small FoV and the time offset between the science and the sky exposures. This yields sky background residuals in the NFM data, which are, on average across the FoV, $\approx25\times$ higher than in WFM data, as well as the presence of skylines that are difficult to be removed.}. To this end, we applied additional processing steps using {\sc CubEx}, following the approach adopted in previous studies of quasar nebulae at similar redshifts \citep[e.g.][]{Borisova+2016, Arrigoni-Battaia+2019, Cantalupo+2019, Fossati+2021}. The first step consists of subtracting the quasar PSF with the {\sc CubePSFSub} routine in {\sc CubEx}. This tool constructs a narrowband image centered on the quasar. From this, a PSF image is generated and rescaled to match the quasar specific flux at each channel within the central $1\times1\,{\rm arcsec}^2$ and $0.25\times0.25\,{\rm arcsec}^2$ region for the WFM and NFM observations, respectively. To mitigate the effect of artifacts and cosmic rays, we compute the rescaling factor using an iterative sigma-clipped mean with a symmetric $\pm3\sigma$ rejection threshold around the median, iterated until convergence. Then, a circular cutout of radius 25 and 50 pixels for the WFM and NFM data, respectively (i.e. $5\arcsec$ and $1.25\arcsec$, corresponding to $\approx 5\times$ and $\approx10\times$ the PSF FWHM) is extracted from the rescaled PSF image and subtracted from the datacube. This procedure is performed for each wavelength layer of the cube. We first ran {\sc CubePSFSub}, setting the spectral filter width ({\tt zPSFsize}) to 600 layers to produce a PSF-subtracted datacube that we visually inspected to identify the spectral regions containing the line emission of the nebulae. We then ran the routine on the original datacube applying masks to such regions and using ${\tt zPSFsize}=150$, corresponding to $187.5\rm\,\AA$. We estimated the FWHM of the PSF of the central quasar by computing a mean radial profile, obtaining $\approx0\rlap{.}{\arcsec}9$ and $\approx0\rlap{.}{\arcsec}12$ in the WFM and NFM, respectively. After the PSF subtraction, we subtract the background and any continuum emission using the {\sc CubeBKGSub} tool \citep[see][]{Borisova+2016}. Briefly, the spectrum of each spaxel is rebinned into spectral bins and median-filtered to estimate the continuum, which is then subtracted from the data. 

\begin{figure*}[!t]
    \centering
    \includegraphics[width=0.6\linewidth]{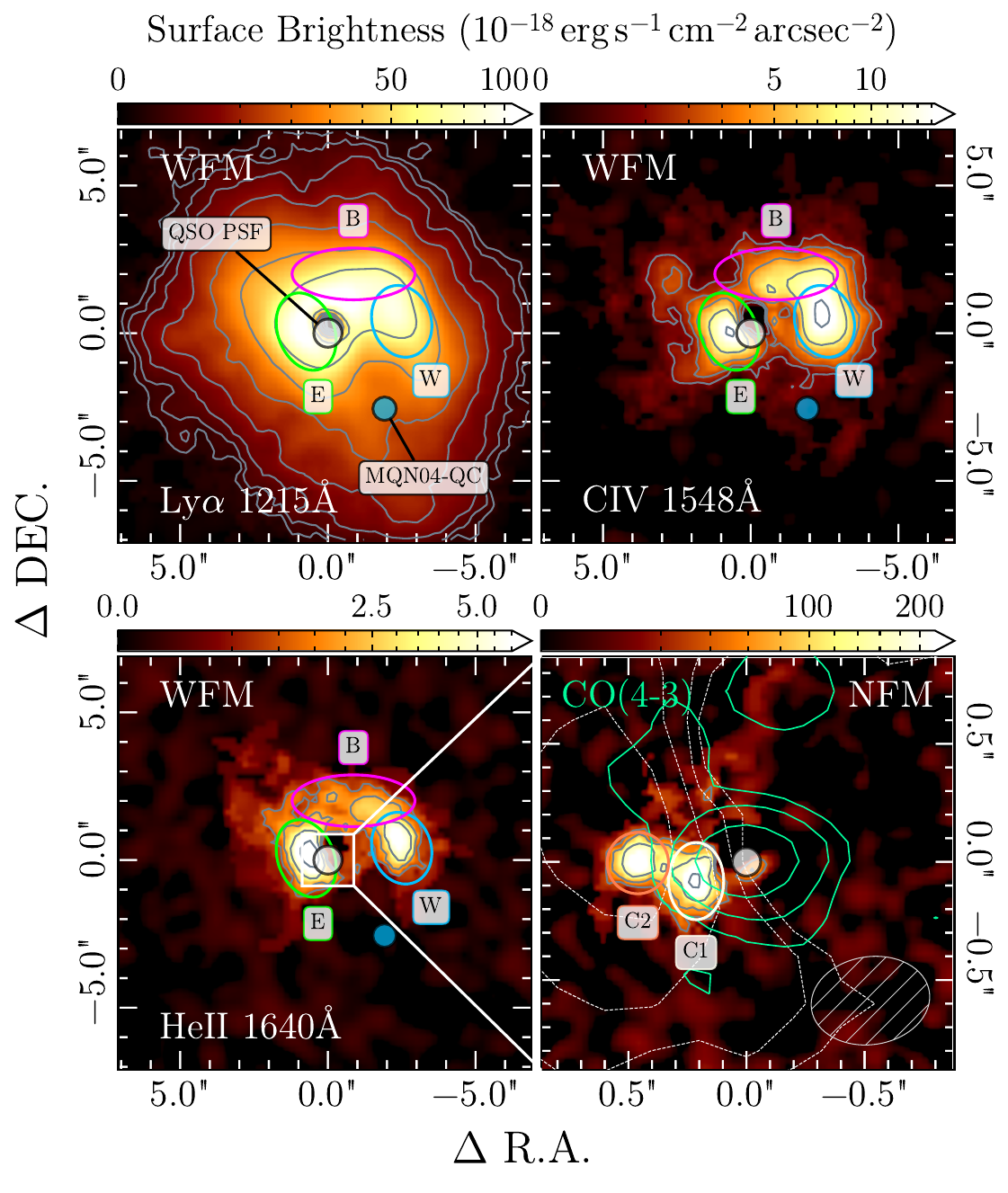}
    \caption{Optimally-extracted images of the extended \lya\ $\lambda1215\,\AA$ (top-left panel), \civ\ $\lambda1548\,\AA$ (top-right panel), and \heii\ $\lambda1640\,\AA$ (MUSE-WFM and NFM data, bottom-left and bottom-right panels, respectively) nebulae detected around the quasar Q0055-269 in the MQN04 field. The contours are matched in flux density and indicate $2^n\times s_0$ SB levels with $n\ge1$ integer, and $s_0 = 0.5\times10^{-18}$ and $16\times10^{-18}\,{\rm erg\,s^{-1}\,cm^{-2}\,arcsec^{-2}}$ for the WFM and NFM, respectively. The layer of the datacube at the quasar redshift has been added as a reference noise background for each transition. The green, magenta, and blue ellipses delineate the selected E (east), B (bridge), and W (west) regions identified based on the \heii\ emission detected in the MUSE-WFM data, while the white and orange ellipses enclose the two clumps (C1 and C2, respectively) revealed by the MUSE-NFM observations. The gray circle indicates the location of the quasar host as determined by ALMA observations, and the size is rescaled to match the PSF FWHM. The blue circle marks the position of MQN04-QC. The contours of the \heii\ nebula detected in the MUSE-WFM data (white), the quasar CO(4--3) emission (green, $2^n\times {\rm RMS}$, with $n\ge1$), as well as the ALMA synthesized beam, are overlaid on the MUSE-NFM data in the bottom-right panel.}
    \label{fig:opt-im} 
\end{figure*}

We then produced optimally-extracted images of the \lya, \civ, and \heii\ emission around the quasar from the WFM data and of the \heii\ emission from the NFM ones. The steps we performed are the following: ({\it i}) we smoothed each layer of the data and of the variance cubes using a two-dimensional Gaussian kernel with standard deviations $\sigma_{x}=\sigma_{y}=1.5\,{\rm pixel}$ (corresponding to $0\rlap{.}{\arcsec}3$ and $0\rlap{.}{\arcsec}04$ for the WFM and NFM data, respectively). ({\it ii}) We searched for all connected voxels above a signal-to-noise threshold of ${\rm S/N}=3.5$ using a friends-of-friends algorithm with a connectivity of 26. We set the minimum size of a structure to be extracted to 2000 voxels. ({\it iii}) We collapsed the datacube by adding together all the selected voxels along the spectral axes. We note that the results are robust against small variations in the adopted parameters. In Fig.~\ref{fig:opt-im} we report the results.

In the WFM datacube we detect bright \lya, \civ, \heii\ emission with ${\rm SB}\ge10^{-18.3}\,{\rm erg\,s^{-1}\,cm^{-2}\,arcsec^{-2}}$. The Ly$\alpha$ emission is the most extended and appears elongated on scales of $\approx130\,{\rm kpc}$ (or $\approx600$~comoving kpc; ckpc) while the \civ\ and \heii\ nebulae have sizes of $\approx55$ and $\approx 45\,{\rm kpc}$ (or $\approx250$, and $210\,{\rm ckpc}$), respectively, as measured at surface brightness levels of ${\rm SB}=10^{-18.3}\,{\rm erg\,s^{-1}\,cm^{-2}\,arcsec^{-2}}$. In the NFM data, we find bright and clumpy \heii\ emission extending eastward up to projected distances of $\approx3.7\rm\,kpc$ (or $\approx20\rm\,ckpc$) from the quasar, which was entirely hidden under the quasar PSF in the WFM observations.

\subsection{The gas in absorption}
\label{ssec:nebule_HI}

To obtain a more complete view of the gaseous environment around Q0055-269, we complemented the emission-line analysis with a search for absorption features in the quasar spectrum obtained with UVES at the redshift of MQN04. As a first step, we converted the wavelength of the UVES spectrum of the quasar from air to vacuum and then visually inspected the continuum-normalized spectrum searching for \hi\ absorption as well as proximate \civ\ and \siv\ absorption-line systems. As previously reported by \citet{Murphy+2019}, no damped \lya\ (DLA) absorbers are present along the quasar sightline. However, we identified an \hi\ absorption system on the blue side of the quasar \lya\ emission showing Ly$\beta ~\lambda1025$\AA\ and \lya\ $\lambda1216$\AA. We did not detect any metal lines associated with this absorber. We fitted Voigt profiles to the Ly$\beta$ and \lya\ features simultaneously to mitigate the uncertainties arising due to the saturation of the latter. To do so, we used the Bayesian code MC-ALF \citep{Longobardi+2023}, following the approach of \citet{Lofthouse+2023} and \citet{Galbiati+2023}. The free parameters are (i) Doppler parameter ($b$); (ii) column density ($N_{\rm\hi}$); (iii) redshift; (iv) a multiplicative constant ranging from 0.98 to 1.02 to account for small uncertainties in the continuum normalization; (v) spectral resolution, allowed to vary within $\pm1\rm\,km\,s^{-1}$ around the nominal value ($\approx7\rm\,km\,s^{-1}$ on average). From this analysis, we determined that the \hi\ absorber lies at $z\approx3.654$ (corresponding to $\Delta\varv_{\rm QSO}\approx-590~{\rm km\,s^{-1}}$ relative to the quasar along the line of sight), has a column density of $\log\,(N_{\rm\hi}/{\rm cm^{-2}})=14.64\pm0.10$ and a Doppler parameter of $b=25.05\pm0.58\rm\,km\,s^{-1}$. The \hi\ column density is much lower than that expected in the self-shielding limit ($\log\,(N_{\rm\hi}/{\rm cm^{-2}})\approx17.2$,\, see, e.g. \citealt{Rahmati+2013}), which indicates that the detected absorber is associated to a gaseous medium that is highly ionized. In Fig.~\ref{fig:spectra}, we report the \hi\ absorber identified in the quasar spectrum along with the best-fit model. We note that additional absorption features are present next to the \hi\ absorber. However, the lack of additional lines of the Lyman series associated with these features prevents us from unambiguously identifying them as \hi.

\subsection{3D multiphase gaseous structure around the quasar}
\label{ssec:nebule_spec}

While \lya\ and \civ\ are the brightest UV/optical emission lines often detected in nebulae around quasars, the non-resonant nature of the \heii\ line emission provides us with a key tracer of the intrinsic properties of the extended, diffuse, ionized gas around the quasar. We produced two-dimensional surface brightness maps by collapsing the datacube around the \lya, \civ, and \heii\ emission lines as described in Sect.~\ref{ssec:nebule_cubex}. In the maps obtained from the MUSE-WFM observations, the \heii\ emission appears to be distributed in two structures, respectively located on the east (E) and west (W) sides with respect to the quasar location, as well as one diffuse emission "bridge" connecting these two regions (B), as shown in Fig.~\ref{fig:opt-im}. Similar morphological features can also be recognized in the \civ\ nebula, and tentatively in the \lya\ map. However, the resonant scattering of the \lya\ photons severely affects the observed spatial distribution of the extended emission, making these structures less apparent. Similarly, we extracted an optimal image of the \heii\ emission from the MUSE-NFM data, which provides a view of the inner $\lesssim5\,{\rm kpc}$ around the quasar. We identified two clumps (C1 and C2) which are located near, but with a measurable displacement, the peak of the \heii\ emission in the E region. Such spatial separation between E and C1/C2 suggests that the emission on large scales is dominated by a diffuse gas component rather than the clumpy structures resolved on kpc scales in the NFM data.

We quantify the velocity offsets among the various regions and characterize the profile of the emission lines by extracting the spectra from the elliptical apertures that encompass the main morphological features identified in the \heii\ nebulae and represented in Fig.~\ref{fig:opt-im}. In Fig.~\ref{fig:spectra} we show the ${\rm S/N}$ spectra of the \lya, \civ\ and \heii\ line emission from the E, W, and B regions as well as those of the \heii\ emission extracted from the C1 and C2 clumps. The emission lines are robustly detected at ${\rm S/N}\gtrsim5-8$ in all these regions, whereas region E suffers from increased noise as a result of its proximity to the quasar PSF. We derived the characteristics of the \heii\ emission in each region identified in the WFM observations by performing a fit of the line profile with a single Gaussian using {\tt emcee} with 50 walkers, and 2500 steps, discarding the first 250 as a burning phase. We report the results in Table~\ref{tbl:heii_lines}. The \heii\ profiles exhibit a similar line width of ${\rm FWHM}\approx 500\,{\rm km\,s^{-1}}$ throughout all three emitting regions. In addition, the view provided by the observation of the gas in emission is complemented by the \hi\ detected in absorption in the quasar sightline (see Sect.~\ref{ssec:nebule_HI}). We found that the \hi\ absorber lies $\lesssim100\,{\rm km\,s^{-1}}$ from the W emitting region. However, we note that the low-column density of this absorber suggests that the observed \hi\ gas is likely arising from the IGM rather than tracing local neutral gas co-existing with the ionized emission. Therefore, the redshift alignment with the W region could plausibly arise by chance. In Sect.~\ref{sec:disc}, we discuss how different scenarios can account for the spatial and kinematical structures revealed by the observed \heii\ emission.

The two bright clumps identified in the NFM observations provide us with additional information on the E region and its connection with the immediate surroundings of the quasar. The \heii\ emission lines extracted from C1 and C2 are broad (${\rm FWHM}\gtrsim500{\rm\,km\,s^{-1}}$) and show a double peaked profile with a minimum within $\lesssim50{\rm\,km\,s^{-1}}$ of the redshift of the quasar. When comparing these \heii\ line profiles with that of the E region extracted from the WFM data, we find that the line centroid is shifted by $\approx-200\,{\rm km\,s^{-1}}$ with respect to the ones measured in C1 and C2. Interestingly, the \heii\ line luminosity in E exceeds by $\approx60\%$ the total luminosity in the combined C1+C2 regions (see Table~\ref{tbl:heii_lines}). Part of the observed difference may be driven by the different PSFs and surface brightness sensitivities of the WFM and NFM observations (see Sect.~\ref{ssec:data_muse}), which affect the apparent flux distribution. However, if the excess flux observed in region E were produced by additional compact clumps, these would be expected to appear in the NFM data with luminosities comparable to those of C1 and C2. The absence of such compact structures in the NFM observations therefore suggests that a significant fraction of the \heii\ emission in E arises from a diffuse gas component on scales larger than $\approx 0\rlap{.}{\arcsec}3$ (or $\approx2.2\,{\rm kpc}$ at $z=3.66$).

\begin{figure*}[!ht]
    \centering
    \includegraphics[width=\linewidth]{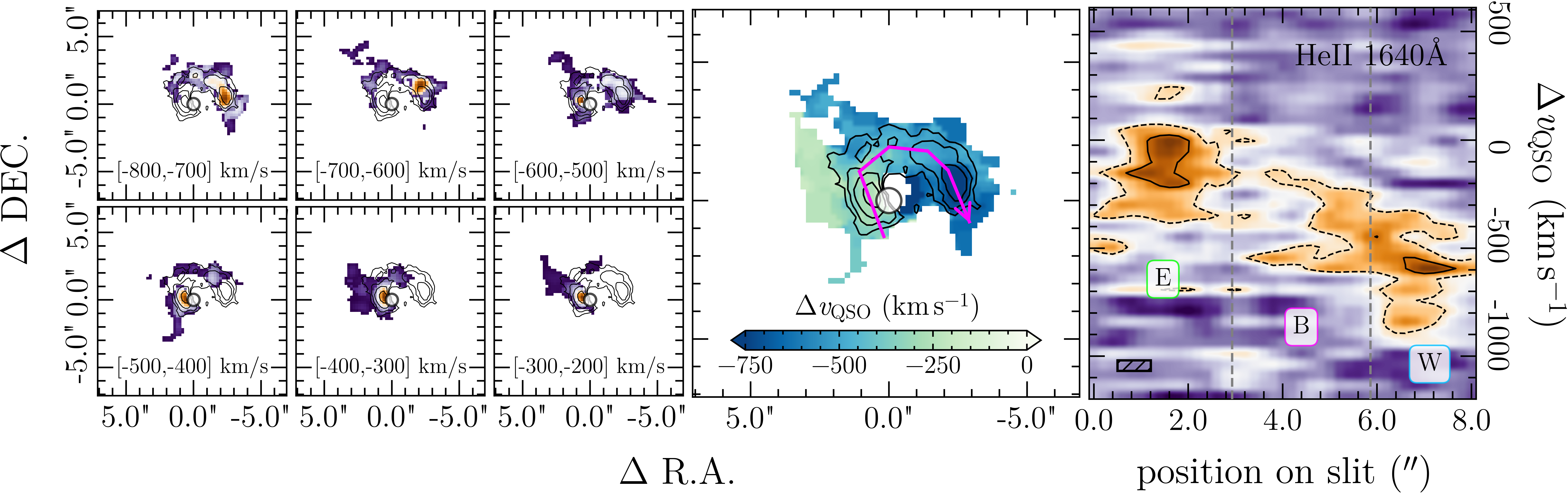}
    \caption{Kinematic analysis of the \heii\ extended emission extracted from the MUSE-WFM data. {\it Left panels}: channel maps of the \heii\ emission obtained from the PSF and continuum-subtracted datacube by collapsing the voxels within windows of $100\rm\,km\,s^{-1}$. The contours refer to \heii\ SB levels as in Fig.~\ref{fig:opt-im}. {\it Central panel}: line-of-sight velocity map (1st moment) of the \heii\ emission. The position of the quasar is indicated by the gray circle with a size corresponding to the QSO PSF FWHM. {\it Right panel}: position-velocity diagram extracted along the polygonal slit shown in the central panel. The boundaries of the E, B, and W region crossed by the slit are reported with vertical lines. The dotted and solid contours indicate $S/N=4$ and $7$ levels, respectively. The spectral and spatial resolution is represented as a rectangle in the bottom-left corner.}
    \label{fig:chan_maps}
\end{figure*}

We then use the non-resonant \heii\ line emission detected in the WFM and NFM data to investigate the 3D structures and kinematics of the gas surrounding the quasar. A kinematical analysis of the MQN04 nebulae was previously performed by \citet{Guo+2020}, although limited by the lack of information on the quasar systemic redshift. In the present work, thanks to our ALMA CO(4--3) line observations (see Sect.~\ref{sec:galaxies}) and the high-resolution MUSE-NFM data, we can now accurately study, for the first time, the observed CGM gas kinematics down to $\approx1\,{\rm kpc}$ relative to the location of the central quasar host galaxy. In Fig.~\ref{fig:chan_maps} we show the channel maps of the \heii\ line emission, the velocity shift relative to the quasar systemic redshift (1st moment), as well as a position-velocity diagram (PVD) extracted from the S/N cube along a polygonal path of 1\arcsec width covering the brightest emitting regions. The two-dimensional maps of the velocity shift and velocity dispersion (2nd moment) are instead shown in Fig.~\ref{fig:heii_moms} and described in App.~\ref{app:heii_moms}. The E and W regions are clearly identifiable not only in the SB maps but also in velocity space, as well as the diffuse gas bridge, labeled B, connecting them. Remarkably, in all these regions, the \heii\ emission appears to be blueshifted with respect to the quasar systemic redshift. Specifically, the \heii\ arising from the W region spans velocities ranging in the interval $\approx[-800,-400]\,{\rm km\,s^{-1}}$, with the peak of the observed emission located at sky-projected distances of $\approx 18\,{\rm kpc}$. The range of observed line-of-sight velocities in this region is also consistent with that of the \hi\ absorber detected along the quasar sightline. This could be indicative that the emitting gas observed in W is part of a more extended structure residing in the foreground of the quasar. On the other hand, in the E region, the \heii\ appears to be closer to the quasar both in projection ($5\,{\rm kpc}$) and in velocity ($\approx[-400,0]\,{\rm km\,s^{-1}}$). This may suggest that at least part of the observed emission arises from the interstellar medium (ISM) of the quasar host galaxy. However, the 1st moment map reveals that the W and E regions appear to be kinematically linked by a smooth velocity gradient. These conclusions are further supported by the inspection of the PVD and the channel maps. Altogether, the combined spatial and kinematical properties of the system suggest that the observed emission traces a physically connected structure rather than multiple unrelated structures. Finally, we note the presence of a low-${\rm S/N}$ emission extending further eastward to and aligned in redshift with the E region, which could instead represent an additional distinct component. 

A similar analysis conducted on lines other than the \heii\ reveals that the gas kinematics traced by the \lya\ and \civ\ emission is more complex and does not show a clear velocity gradient across the nebulae. However, except for the E region, the overall velocity shifts are consistent with those measured from the \heii\ line (see Fig.~\ref{fig:spectra}). This indicates that in B and W regions radiative transfer effects seem to only moderately affect the ability of resonant lines to trace the gas kinematics, as expected in a highly ionized medium.

\begin{table}[!t]
\def\arraystretch{1.15}
\caption{Properties of the HeII emission and line flux ratios.}  
\label{tbl:heii_lines}    
\centering
\resizebox{1\hsize}{!}{
\begin{tabular}{l c c c}
\toprule
\toprule
WFM & E region & W region & B region \\
\cmidrule(lr){1-4}
$R_{\rm sky, \, QSO}\,^{(\dagger)}$ (kpc)                 &$5.4$      &$18.7$     &$16.1$ \\
$\Delta\varv_{\,{\rm \heii},\,{\rm QSO}}$ (km s$^{-1}$)  & $-139\pm18$ & $-587\pm13$ & $-414\pm15$\\
${\rm FWHM}_{\,\rm \heii}$ (km s$^{-1}$)                & $579\pm47$ & $486\pm23$ & $501\pm32$\\
$\log[L_{\rm\heii}/({\rm erg\,s^{-1}})]$ & $42.51\pm0.03$ & $42.27\pm0.02$ & $42.20\pm0.32$ \\ 
\civ/\lya  & $0.086\pm 0.005$ & $0.163 \pm 0.003$ & $0.099 \pm 0.003$\\
\civ/\heii & $1.47\pm 0.09$ & $3.69\pm 0.19$ & $4.5\pm0.5$ \\
\heii/\lya & $0.074\pm 0.007$ & $0.0471\pm 0.0011$ & $0.0333\pm0.0011$\\
\cmidrule(lr){1-4}
NFM & C1 region & C2 region & \\
\cmidrule(lr){1-4}
$R_{\rm sky, \, QSO}\,^{(\dagger)}$ (kpc)                 &$1.5$      &$3.7$     & \\
$\Delta\varv_{\,{\rm \heii},\,{\rm QSO}}$ (km s$^{-1}$)  & $63\pm15$ & $20\pm14$ &\\
${\rm FWHM}_{\,\rm \heii}$ (km s$^{-1}$)                & $530\pm40$ & $508\pm28$ &\\
$\log[L_{\rm\heii}/({\rm erg\,s^{-1}})]$ & $41.80\pm0.03$ & $41.76\pm0.03$ & \\
\bottomrule
\end{tabular}
}
\tablefoot{$^{(\dagger)}$Sky-projected distance of the centroid of the region from the quasar location measured from the \heii\ optimally-extracted surface brightness map (see Fig.~\ref{fig:opt-im}).}
\end{table}

\subsection{Spatially resolved line ratios}
\label{ssec:line_ratios}

\begin{figure*}[!t]
    \centering
    \includegraphics[width=0.85\linewidth]{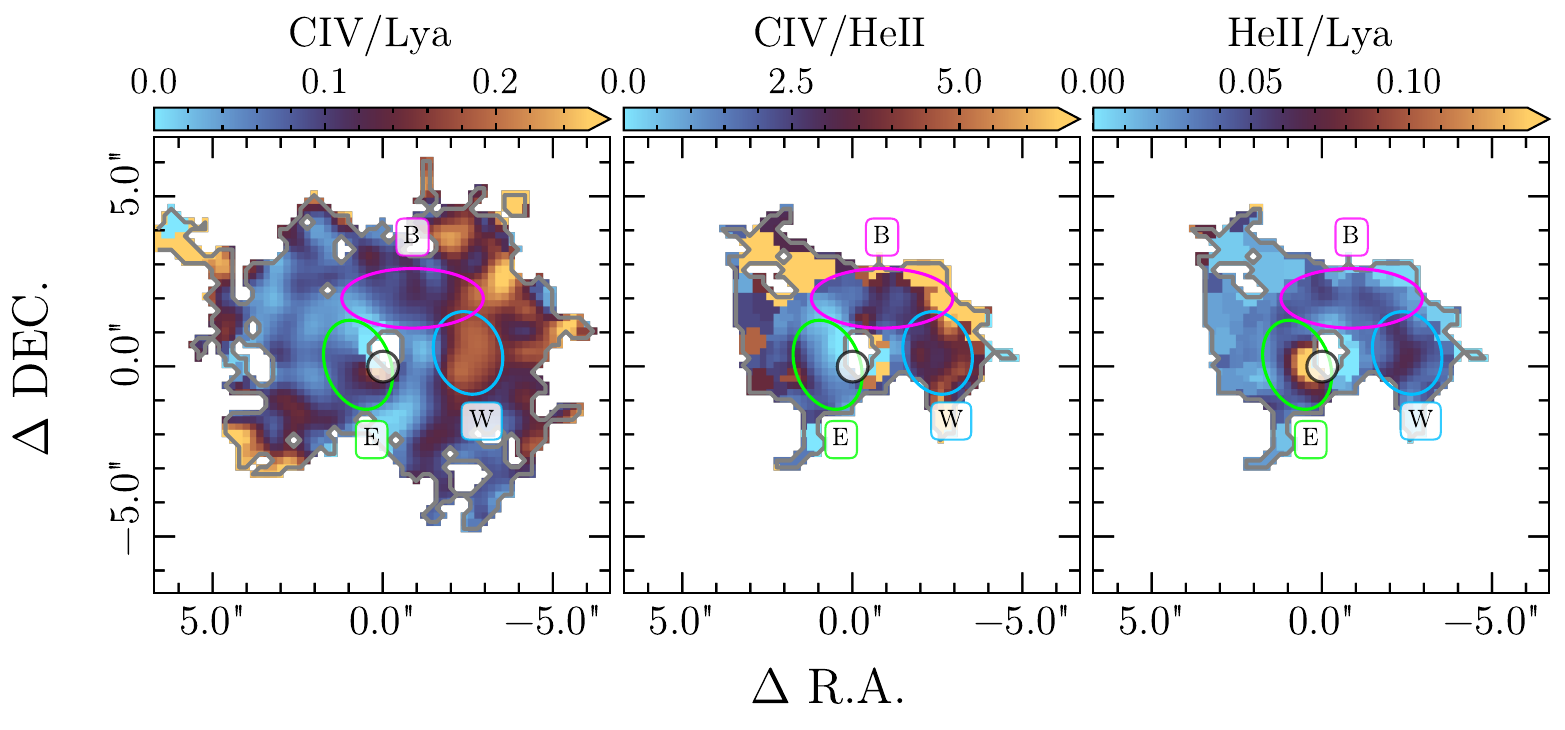}
    \caption{Line ratios in the MQN04 nebula: \civ/\lya\ (left panel), \civ/\heii\ (central panel), \heii/\lya\ (right panel). The line ratios are computed in Voronoi-resampled datacubes to maximize the signal-to-noise of the faintest regions. The gray contour in each panel represents the SB mask of the optimally extracted image, as in Fig.~\ref{fig:opt-im}, of the least extended nebula included in the line ratio. The colored ellipses mark the same regions identified in Fig.~\ref{fig:opt-im}. The gray circle indicates the location of the quasar host as determined by ALMA observations and is sized as the PSF FWHM.}
    \label{fig:line-ratios}
\end{figure*}

The simultaneous detections of \lya, \civ, and \heii\ emission lines provide powerful diagnostics to get insights into the origin and physical properties of the extended gas nebulae surrounding the quasar such as the ionization parameter, gas density and metallicity \citep[e.g.][]{Villar-Martin+1997, ArrigoniBattaia+2015a, ArrigoniBattaia+2015b, Prescott+2015, Cantalupo+2019, Marino+2019, Jimenez-Andrade+2023}. Thanks to the exquisite S/N of the observations presented in this work, we can obtain spatially resolved line ratios in the MQN04 nebula. To this purpose, we maximize the recovery of faint \heii\ and \civ\ emission by applying a resampling scheme along the spatial axes of the smoothed datacube using the adaptive two-dimensional Voronoi binning technique \citep{Cappellari+2003}. In this approach, we define each cell to achieve ${\rm S/N}=8,9$ thresholds, for the \heii\ and \civ, respectively, chosen to preserve the native sampling in the regions of interest while maximizing the detection of faint emission in low S/N pixels. To obtain consistent and optimal line ratio maps, we resample the datacubes using the Voronoi binning obtained for the least extended nebula of the ratios. We report the results in Fig.~\ref{fig:line-ratios} and in Table~\ref{tbl:heii_lines}. 

Interestingly, all the line ratios are significantly different between the E and W regions, indicating that different physical properties and/or ionization conditions are in place. In particular, following \citet{Cantalupo+2019}, by assuming that the observed emission is powered by photoionization and the medium is characterized by a single density distribution, the observed difference between the \heii/\lya\ ratios indicates that the E and W regions are located at different physical distances. Under the same hypotheses, the \heii\ emission in the E region is enhanced due to its proximity to the quasar ionization source. Furthermore, the high \civ/\lya\ and \civ/\heii\ line ratio observed in W region may indicate significant metal enrichment of the gas \cite[e.g.][]{ArrigoniBattaia+2015b, Marques-Chaves+2019, Kolwa+2019}. However, the determination of the metallicity of this region is challenging given the lack of any metal absorption features at the redshift of the \hi\ aligned with W in the UVES spectrum of the quasar. In Sect.~\ref{sec:disc}, we discuss various scenarios to interpret these results. 

\subsection{Overdensity of SF galaxies and large-scale environment}
\label{ssec:over_galaxies}

Direct observations of the cosmic gas in emission, particularly in regions surrounding quasars \citep[e.g.][]{Cai+2018, Tornotti+2024} and hosting dense concentrations of star-forming galaxies \citep{Cai+2016, Umehata+2019, Pensabene+2024, Banerjee+2025, Galbiati+2025, Tornotti+2025}, offer powerful means to probe its large-scale distribution and explore its connection to galaxy evolution. We inspected the MUSE data to conduct a census of the population of star-forming galaxies embedded in the gaseous structures observed in emission, as described in Sect.~\ref{sec:galaxies}. We identified 5 galaxies with spectroscopic redshifts within $\abs{\Delta\varv_{\rm QSO}}\lesssim500\rm\,km\,s^{-1}$ (median $\Delta\varv_{\rm QSO}\approx-68\rm\,km\,s^{-1}$) from the systemic redshift of the central quasar. As shown in Fig.~\ref{fig:pin-point-gals}, three of these galaxies are concentrated within $\approx500\rm\,ckpc$ and embedded in the \lya\ nebula. The remaining two, although lying at larger projected distances from the quasar, appear to be aligned along the direction in which the nebula extends. This suggests that filaments of gas may continue further along this axis and may not be detected in emission in the current data due to their lower density or because they are not efficiently illuminated by the radiation of the quasar or additional undetected AGNs. 

We are interested in comparing this system with other known overdensities found at similar redshifts and proven to be connected by filaments of gas observed in \lya\ emission. We compared MQN04 field with two other structures for which MUSE data with similar exposure times are available, namely the protoclusters SSA22 at $z\approx3.08$ \citep{Umehata+2019} and MQN01 at $z\approx3.25$ \citep{Pensabene+2024, Galbiati+2025}. For these comparison fields, we repeated the same procedure described in Sect.~\ref{ssec:galaxies_muse} and selected all the star-forming galaxies with R-band magnitudes $m_{\rm R}\le27\rm\,mag$, which is the completeness limit of the samples based on the luminosity functions in \citet{Galbiati+2025}. We then measured the number density of these galaxies within velocities $\abs{\Delta\varv}\le1000\rm\,km\,s^{-1}$ and projected separation $R\le20\arcsec$ (defined to match the FoV of MQN04) from the central quasar in MQN01 field. In the case of SSA22, since there are no bright quasars in the structure, we took as the fiducial value of the number density the median of the distribution obtained by bootstrapping over the position and redshift of the individual galaxies, taken one by one as the center of the structure. We chose the same velocity interval as MQN01, i.e. $\abs{\Delta\varv}\le1000\rm\,km\,s^{-1}$, and projected separation $R\le2\arcmin$, calibrated to encompass at least five neighbors since the galaxies are more sparse in the FoV. As a control field sample, we took that described by \citet{Galbiati+2025} assembled from fields where quasars are not present at $z\approx3$ that were observed with MUSE and have depth similar to MQN04 data.

As a result, we measure the following comoving number densities: $\rho_{\rm MQN04}=(0.041\pm0.002)\rm\,cMpc^{-3}$, $\rho_{\rm MQN01}=(0.056\pm0.007)\rm\,cMpc^{-3}$, $\rho_{\rm SSA22}=(0.015\pm0.002)\rm\,cMpc^{-3}$, and $\rho_{\rm field}=(0.0011\pm0.0001)\rm\,cMpc^{-3}$. This implies overdensities relative to the field of $\delta_{\rm MQN04}=41\pm4$, $\delta_{\rm SSA22}=15\pm2$, and $\delta_{\rm MQN01}=56\pm9$. We note, however, that the larger projected radius adopted for SSA22 results in a significantly larger surveyed volume compared to the other structures. This difference may dilute the measured galaxy number density if the structure is centrally concentrated, and therefore, the overdensities should be compared with this caveat in mind. We found that MQN04 is substantially more overdense than SSA22, though not as extreme as MQN01. In MQN04 field, the galaxies appear spatially concentrated around the quasar and lie within small velocity separations. However, none display AGN-like emission lines (e.g. \ion{N}{V} $\lambda1240$\AA, \ion{C}{IV} $\lambda1548,1550$\AA, \ion{He}{II} $\lambda1640$\AA). Assuming the same AGN fraction observed in MQN01 for galaxies with $M_\star\gtrsim10^{9.5}\,M_\odot$ ($f_{\rm AGN}\approx0.4$; \citealt{Travascio+2025}), we would expect roughly two AGN in MQN04, yet none are detected. Furthermore, the only SMG detected with ALMA, MQN04-QC, lies relatively far from the quasar in both redshift and projected separation. This contrasts with MQN01, where a significant AGN overdensity is clustered within an area of $2\arcmin\times2\arcmin$ around the central quasar \citep{Travascio+2025}, and a massive submillimeter companion lies only $\Delta\varv\approx-300\rm\,km\,s^{-1}$ and $10\rm\,kpc$ from the quasar \citep{Pensabene+2025}. Interestingly, none of these features is present in SSA22.

Overall, the three structures appear to trace different environments and may occupy distinct locations within the cosmic web. SSA22 is relatively diffuse, MQN04 is denser yet still AGN-poor, and MQN01 is highly overdense and AGN-rich, with the latter showing the strongest indication of tracing a massive node (Cantalupo et al., in prep.). Alternatively, the differences among them may reflect observations of structures at different evolutionary stages \citep[e.g.][]{Shimakawa+2018b}.

\section{Discussion}
\label{sec:disc}

The extended gas directly imaged via multiple emission lines in the MQN04 field provides powerful insights into the complex and multiphase structure of the CGM gas. In particular, the morphology of the non-resonant \heii\ emission suggests the presence of multiple components that could be spatially and kinematically connected. In what follows, we address the plausibility of three different scenarios as possible explanations for the observed characteristics of the diffuse gas.

\subsection{Scenario A: inflow of a gas stream}
\label{ssec:disc_inflow}

We first investigate the possibility that the observed \heii\ line emission arises from inflows of cool gas streams that accrete onto the quasar host galaxy from intergalactic scales, as predicted by hydrodynamical simulations \citep[e.g.][]{Fardal+2001, Keres+2005, Dekel+2009, vandeVoort+2012, Waterval+2025}. Since a large portion of the gas within these streams can exhibit high angular momentum, the accreting gas may appear as an inspiraling structure \citep[e.g.][]{Danovich+2015, Stewart+2017, Arrigoni-Battaia+2018, WangS+2022, Zhang+2023b, Zhang+2023a} and produce velocity gradients as observed in MQN04 (see Fig.~\ref{fig:chan_maps}). 

In this framework, the velocity structure of the gas can be interpreted as follows. The blueshifted region W would be a stream of gas infalling toward the quasar host from the far side, while the lower line-of-sight velocity in the bridge B would be due to a change in direction as a result of the inspiraling motion. Finally, region E and clumps C1/C2, close to the QSO systemic redshift, could represent the last leg of the accretion path, where the inflowing gas merges with the ISM of the central galaxy. We note that diffuse blueshifted emission further eastwards of region E cannot be explained by the same stream originating in region W and should rather be attributed to a separate accretion flow. Within this scenario, the absence of any extended redshifted emission would suggest that neither stream makes a full turn around the galaxy, indicative either of relatively low angular momentum, or that some angular momentum has been lost during the inflow, possibly due to interactions with the hot phase of the CGM \citep[see, e.g. the semi-analytical models from][]{Afruni+2021, Lan+2019}. In the scenario described above, the \hi\ absorption could be seen as a \lya\ forest absorber that is physically unrelated to the emitting gas and appears aligned with W in velocity just for coincidence.

Instead, if the emitting gas in W is physically associated with the \hi\ absorber, as similar velocities suggest, then it must be located on the near side. The observed blueshift can be reconciled with an inflow scenario only if the gas extends over intergalactic scales where the Hubble flow dominates its relative motion. In this case, assuming that the inflow velocity of the gas is small compared to the Hubble flow, the inflow should be a contiguous filament of gas extending at least $\approx1.6\rm\,Mpc$ mainly aligned along the line-of-sight. However, the observed high surface brightness of the \heii\ emission cannot be explained at such large distances from the quasar without the presence of additional multiple AGN with a hard ionizing spectrum \citep[e.g.][]{Humprey+2019, Pezzulli+2019}. While a possible source of such ionizing radiation may be identified in MQN04-QC, assuming it hosts an obscured AGN, it still lies too far away both in redshift space and in projection from the \heii\ emitting gas. Also, no evidence for additional AGN appears in the data. This scenario would also require the presence of some redshifted emission arising from distances smaller than the turnaround radius \citep[$\approx4~R_{\rm vir}$, e.g.,][]{deBeer+2023}, where the inflow velocity dominates over the Hubble flow, which is not observed. While several aspects of these scenarios cannot be confirmed with the existing observations, and despite the limitations discussed above, the inflow interpretation remains a plausible explanation for the observed properties.

\subsection{Scenario B: multiphase ionized outflow}
\label{ssec:disc_outflow}

Another interpretation for the blueshifted emission and \hi\ absorption could be an outflow. A \emph{standard} axisymmetric biconical outflow scenario can be ruled out, unless the redshifted components are completely obscured by dust \citep[e.g.][]{Villar-Martin+2011}. However, there is no indication from our deep ALMA observations of the presence of dust on CGM scales. An outflow with more complex geometry (non-biconical) cannot be excluded. We note that the spatial variations in the \heii/\lya\ line ratios (see Sect.~\ref{ssec:line_ratios}) could indicate that E and W (as well as B) regions are located at different physical distances from the quasar. This pattern could arise if outflows propagate through media of unequal resistance, caused either by the dynamical pressure of denser gas, such as cold inflows, or by turbulent viscosity \citep[e.g.][]{Gabor+2014}. Alternatively, the observed regions might trace a single outflow whose cones have varying orientations or were launched at different epochs. In addition, a hint of a blueshifted excess in the CO(4--3) emission line profile of the quasar host at $\approx-300\,{\rm km\,s^{-1}}$ (see Fig.~\ref{fig:spectra}) might be associated with the molecular phase of an outflow. However, we do not observe CO(4--3) emission co-spatial with the extended \heii\ emission. This might suggest that molecules are dissociated within the outflow on large scales. Deep ALMA data combined with IFS observations with JWST/NIRSpec targeting H$\alpha$ emission to map the ionized gas at high resolution may help further explore this scenario.

\subsection{Scenario C: stripping from a galaxy encounter}
\label{ssec:disc_merger}

Alternatively to the previous scenarios, we consider the possibility of tidal stripping due to a close encounter between a gas-rich galaxy and the quasar host. To test this scenario, we compute a lower limit on the quasar halo mass assuming that the close encounter occurs with the following simplifying conditions: (i) the observed velocity of the gas in W, the brightest and extended region, corresponds to the galaxy velocity; (ii) the observed line-of-sight velocity of the gas in W ($\approx -600\,{\rm km\,s^{-1}})$ is entirely due to its peculiar motion relative to the quasar systemic redshift; (iii) the gas is moving with velocity fully aligned with the line of sight, and (iv) it is indicative of the velocity of the galaxy at the pericenter of its orbit. By equating the observed velocity of the galaxy with the escape velocity\footnote{The escape velocity is defined as $\varv_{\rm esc}(R)=\sqrt{-2\phi(R)}$. Here we assume $\phi(R)$ as the gravitational potential of the dark matter halo with a NFM density profile and concentration index $c=4$ from \citealt{Navarro+1996, Rodriguez-Puebla+2016}.}, we obtain that a minimum halo mass of $M_{\rm halo, min}^{200}>8\times10^{11}{\rm\,M_{\astrosun}}$ is required to accelerate a test mass up to $600\,{\rm km\,s^{-1}}$ at the pericenter distance of $\approx18.7\,{\rm kpc}$ (see Table~\ref{tbl:heii_lines}). The estimated lower limit is consistent with the typical halo mass of quasars at cosmic noon \citep[e.g.][]{Trainor+2012, Eftekharzadeh+2015, deBeer+2023, Pizzati+2024}. To complete the picture, the region B and the \hi\ absorber could trace different portions of the tidal tail arising from the quasar-galaxy interaction, while E and the region extending further eastward could represent gas stripped during a previous close passage. 

Alternatively, the large amount of diffuse, extended gas detected around the quasar can also be explained by the presence of a hot medium, such as a proto-intracluster medium (ICM) \citep[e.g.][]{DiMascolo+2023} that is efficient in stripping the gas from a galaxy passing through. The large overdensity of star-forming galaxies clustered around the quasar (see Sect.~\ref{ssec:over_galaxies}) supports the hypothesis of a massive quasar host halo, as detected, for instance, in MQN01 \citep{Travascio+2025}. 

While the galaxy encounter scenario currently offers a plausible explanation for the observed extended \heii\ emission, confirming the existence of galaxies physically interacting with the quasar, and eventually unveiling the hypothesized hot CGM, will require deeper, higher-resolution, multi-wavelength imaging and spectroscopic observations. We emphasize that the scenarios discussed above are not exhaustive: more complex configurations remain possible, including, for example, an asymmetric outflow triggered by a major merger event, or an asymmetric inflow of gas previously stripped from a companion galaxy. We note that all the scenarios described above may need to be revised to account for the possibility that the quasar host does not trace the center of mass of the system and its redshift measurement is affected by its proper motion \citep[e.g.][]{Pensabene+2025}. Finally, it also remains to be understood what is the origin of the more extended portions of the \lya\ nebula currently undetected in \heii.

\section{Summary and Conclusions}
\label{sec:conclusions}

In this work, we presented ALMA and MUSE-NFM observations toward the field of the quasar Q0055-269 at $z\approx3.66$ and the surrounding MQN04, one of the brightest \lya\ nebulae known at high redshift, which is also detected in \civ\ and \heii\ emission extending over $\gtrsim50\rm\,kpc$ \citep{Borisova+2016, Guo+2020}. Combined with the view of the ionized gas phase offered by previous MUSE-WFM observations, this new data allows us to resolve the kinematics, morphology, and clumpiness of the CGM gas flows on scales ranging from tens of kiloparsecs down to $\approx1\rm\,kpc$ away from the quasar.

\begin{itemize}
    \item We detected both the (observed-frame) 3-mm continuum and the CO(4--3) line emission from the quasar host galaxy, providing the first measurement of its systemic redshift and allowing us to accurately study the kinematics of the surrounding ionized gas relative to the quasar. We also detected the CO(4--3) line emission from a second source, MQN04-QC, offset by $\Delta\varv_{\rm QSO}\approx-1100\rm\,km\,s^{-1}$ from the quasar. The non-detection of any stellar continuum in UV observations suggests that MQN04-QC is a heavily dust-obscured galaxy.    
    \item The combined analysis of the \lya, \civ, and \heii\ lines reveals emission extending over $\approx50\rm\,kpc$ in projection. We distinguish two main bright structures on the east and west sides of the quasar host, which appear kinematically connected by a bridge of diffuse gas. The emission line profiles throughout these regions are broad (${\rm FWHM}\gtrsim500\rm\,km\,s^{-1}$) and blueshifted by $\approx0-800\rm\,km\,s^{-1}$ relative to the quasar systemic redshift. Interestingly, the most blueshifted component aligns in velocity with an intervening \hi\ absorber with a column density $N_{\rm\hi}=10^{14.64}\rm\,cm^{-2}$ detected in the quasar spectrum, thus suggesting that the gas is highly ionized. As expected for resonant lines, \lya\ and \civ\ show smoother spatial and velocity distributions than non-resonant \heii\ yet, with the exception of the central region, they do not exhibit a systematic velocity shift relative to \heii. This indicates that these lines reliably trace the average ionized-gas kinematics across most of the nebula, likely because quasar illumination keeps the CGM highly ionized and radiative-transfer effects relatively modest. On smaller scales, the MUSE-NFM observations uncover two additional compact \heii-emitting clumps located $\approx5\rm\,kpc$ away from the quasar and at a similar redshift. Similar to the emission on larger scales, these structures are characterized by broad (${\rm FWHM}\gtrsim500\rm\,km\,s^{-1}$) and complex \heii\ line profiles. We explored three possible scenarios to explain the observed \heii\ emission: an inflowing gas stream, an outflow, and gas stripping induced by a close galaxy encounter or by a passage of a galaxy through a hot CGM. 
    \item We conducted a census of the galaxy population and identified five star-forming galaxies clustered within $\abs{\Delta\varv_{\rm QSO}}\lesssim1000\rm\,km\,s^{-1}$ around the central quasar. Three of them are embedded in the \lya\ nebula, while the remaining two are located at larger distances, but aligned with the orientation of the nebula, and we speculate may also be connected with fainter still undetected \lya\ emission. We measure an overdensity of $\delta\approx41$. Compared to other overdensities at similar redshifts, this structure resembles known protoclusters (e.g. MQN01) but lacks enhanced AGN activity or an excess of submillimeter galaxies. This may indicate that such systems trace different regions of the cosmic web or are captured at distinct stages of their evolution.
\end{itemize}

Follow-up observations with ALMA achieving higher sensitivity and with JWST/NIRSpec will be crucial to map the cold molecular gas distribution and detect additional non-resonant lines, such as H$\alpha$, to uncover both the mechanisms powering the extended emission and any additional AGN or massive quiescent galaxies that remain undetected in the current data. Our joint analysis of spatially-resolved kinematics traced by the non-resonant \heii\ emission and secure systemic redshift determination via CO(4--3) line, show that MQN04 fits within the broader picture in which gas flows from large to small scales are shaped by galaxy interactions and by the interplay between feeding and feedback. Most importantly, for the first time, we unambiguously measure the presence of a strong systematic velocity shift between the CGM kinematics and the quasar host systemic redshift, demonstrating that the CGM in this system is highly asymmetric. Overall, systems like MQN04 are ideal laboratories for studying the baryon cycle in massive halos at high redshift and establishing how the interplay between the kinematics of different gas phases and the large-scale environment drives the coevolution of galaxies and the cosmic web across cosmic time. 


\begin{acknowledgements}
We thank the anonymous referees for their valuable suggestions, which helped improve this work. We thank Mr. Cosimo Marconcini and Dr. Giacomo Venturi for their contribution and insightful discussion on the outflow scenario. This project was supported by Progetto FARE 2020 {\it Svelare i nodi massicci della CosmicWeb} ID 2021-NAZ-0326/PER, the European Research Council (ERC) Consolidator Grant 864361 (CosmicWeb), and by Fondazione Cariplo grant no. 2020-0902. AP acknowledges the support from the Independent Research Fund Denmark (DFF) under grant 3120-00043B. This paper makes use of the following ALMA data: ADS/JAO.ALMA\#2024.1.00499.S ALMA is a partnership of ESO (representing its member states), NSF (USA) and NINS (Japan), together with NRC (Canada), NSTC and ASIAA (Taiwan), and KASI (Republic of Korea), in cooperation with the Republic of Chile. The Joint ALMA Observatory is operated by ESO, AUI/NRAO and NAOJ. This work is also based on observations collected at the European Southern Observatory under ESO programmes IDs  094.A-0131(B), 096.A-0222(A), 109.232M and data obtained from the ESO Science Archive Facility with DOI under \url{https://doi.eso.org/10.18727/archive/42}. This research made use of Astropy\footnote{\url{http://www.astropy.org}}, a community-developed core Python package for Astronomy \citep{AstropyI, AstropyII}, NumPy \citep{Numpy}, SciPy \citep{Scipy}, Matplotlib \citep{Matplotlib}.
\end{acknowledgements}

\bibliographystyle{aa}
\bibliography{MyBib_over_latest}


\begin{appendix}
\onecolumn
\section{MUSE-selected galaxies in the field}
\label{app:muse-gal}

In  Table~\ref{tbl:muse_galaxies}, we list the galaxies detected by MUSE in the MQN04 field within $\abs{\Delta\varv_{\rm QSO}}\le1000\,{\rm km\,s^{-1}}$.

\begin{table}[!t]
\def\arraystretch{1.15}
\caption{Properties of the star-forming galaxies identified in MUSE and clustered within $\abs{\Delta\varv_{\rm QSO}}\le1000\rm\,km\,s^{-1}$ around the systemic redshift of the central quasar.}  
\label{tbl:muse_galaxies}    
\centering
\resizebox{0.65\hsize}{!}{
\begin{tabular}{l c c c c c}
\toprule
\toprule
& ID 1 & ID 2 & ID 3 & ID 4 & ID 5 \\
\midrule
R.A. (ICRS) & 00:57:56.82 & 00:57:57.65 & 00:57:57.87 & 00:57:57.58 & 00:57:58.56 \\
Dec. (ICRS) & -26:43:32.3 & -26:43:24.7 & -26:43:22.5 & -26:43:12.9 & -26:42:57.7 \\
$z_{\rm spec}$ & 3.656 & 3.669 & 3.663 & 3.659 & 3.667 \\
$\Delta\varv_{\rm QSO}\,({\rm km\,s^{-1}})$ & -507 & 357 & -68 & -318 & 173 \\
Spectral features$^{(\dagger)}$ & \lya & \lya & \lya &  \lya & \lya, \ion{O}{I}+\ion{SiII}, \ion{C}{II}  \\
$m_{\rm r}$~(mag) & 26.98$\pm$0.10 & 26.96$\pm$0.12 & 26.41$\pm$0.08 & 26.30$\pm$0.11 & 25.44$\pm$0.07 \\
\bottomrule
\end{tabular}
}
\tablefoot{$^{(\dagger)}$Emission and absorption lines identified in the spectra and used to constrain the spectroscopic redshift.}
\end{table}

\section{Velocity field and velocity distribution of the \heii\ nebula}
\label{app:heii_moms}

\begin{figure}[!t]
    \centering
    \includegraphics[width=0.5\linewidth]{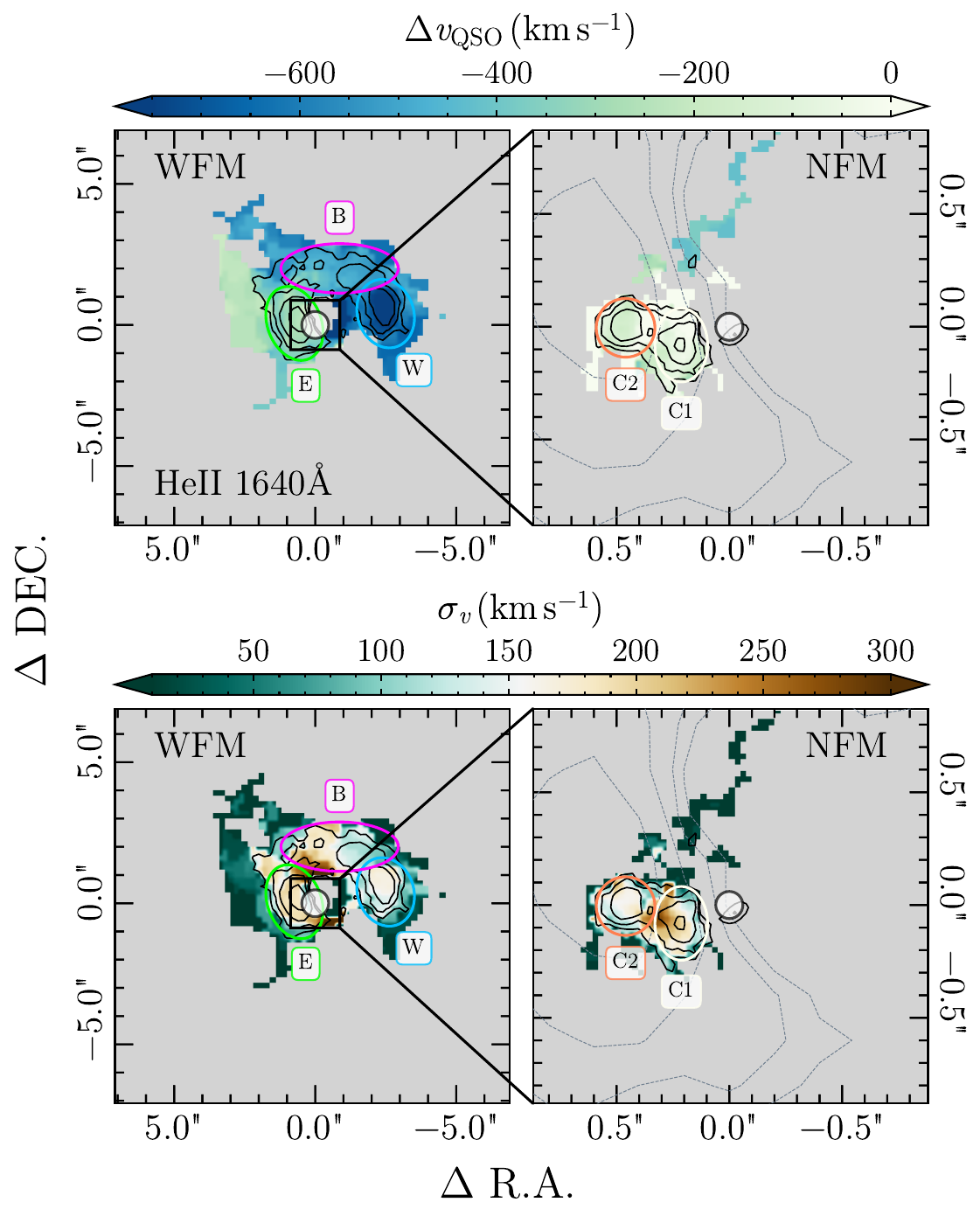}
    \caption{The observed \heii\ gas velocity shift relative to the quasar host galaxy systemic redshift (1st moment, {\it top panels}), and velocity dispersion (2nd moment, {\it bottom panels}) as derived from the MUSE WFM and NFM data ({\it left and right panels}, respectively). The contours refer to the \heii\ SB levels as in Fig.~\ref{fig:opt-im}. The regions identified in Sect.~\ref{ssec:nebule_spec} are also reported.}
    \label{fig:heii_moms}
\end{figure}

In Fig.~\ref{fig:heii_moms}, we present \heii\ velocity shift (1st moment) and dispersion (2nd moment) extracted from MUSE WFM and NFM datacubes. The \heii\ emitting gas appears blueshifted up to $\approx-700\,{\rm km\,s^{-1}}$ with respect to the quasar systemic redshift determined by the ALMA CO(4--3) line detection, differently from what is reported by \citet{Guo+2020} who employed as a reference frame for the line-of-sight velocities the redshift derived from the \lya\ emission line profile of the diffuse nebula. The observed velocity dispersion reaches values of $\gtrsim300\,{\rm km\,s^{-1}}$ around regions E and B, consistent with the findings of \citet{Guo+2020}. In addition, NFM data reveal similarly high velocity dispersion values in the regions C1 and C2, within the inner few kiloparsecs from the quasar host galaxy. Such an increase of the \heii\ gas velocity dispersion in the proximity of the quasar might be associated with higher turbulence caused by merger activity, outflows, or gas inflows into the gravitational potential well of the quasar halo (see Sect.~\ref{sec:disc}).

\end{appendix}

\end{document}